\documentclass{llncs}
\AtBeginDocument{%
  \providecommand\BibTeX{{%
    \normalfont B\kern-0.5em{\scshape i\kern-0.25em b}\kern-0.8em\TeX}}}
\usepackage{graphicx}
\usepackage{multicol}
\usepackage{algorithm}
\usepackage{algorithmic}
\def\B{\mathcal{B}}
\def\C{\mathcal{C}}
\def\D{\mathcal{D}}
\def\E{\mathcal{E}}
\def\G{\mathcal{G}}
\def\M{\mathcal{M}}
\def\N{\mathcal{N}}
\def\F{\mathcal{F}}

\def\U{\mathcal{U}}

\def\dto{\Rightarrow}

\long\def\BEGINOMIT#1\ENDOMIT{\relax}
\def\url#1{{\tt#1}}

\begin{document}

\title{Analysis of Co-Occurrence Patterns in Data through 
Modular and Clan Decompositions of Gaifman Graphs\thanks{%
This research was supported by
{European Research Council} ERC-2014-CoG 648276 (AUTAR), by
TIN2017-89244-R from Ministerio de 
Economia, Industria y Competitividad, and by
Conacyt (M\'exico); 
we acknowledge recognition 2017SGR-856
(MACDA) from AGAUR (Generalitat de Catalunya).%
}}

\author{Jos\'e Luis Balc\'azar\inst{1} \and Marie Ely Piceno\inst{2}} 
\institute{Universitat Polit\`ecnica de Catalunya}


\BEGINOMIT
<ccs2012>
   <concept>
       <concept_id>10003120.10003145.10003147.10010923</concept_id>
       <concept_desc>Human-centered computing~Information visualization</concept_desc>
       <concept_significance>500</concept_significance>
       </concept>
   <concept>
       <concept_id>10002950.10003624.10003633.10010917</concept_id>
       <concept_desc>Mathematics of computing~Graph algorithms</concept_desc>
       <concept_significance>300</concept_significance>
       </concept>
   <concept>
       <concept_id>10003120.10003145.10003146.10010892</concept_id>
       <concept_desc>Human-centered computing~Graph drawings</concept_desc>
       <concept_significance>100</concept_significance>
       </concept>
   <concept>
       <concept_id>10002951.10003227.10003351</concept_id>
       <concept_desc>Information systems~Data mining</concept_desc>
       <concept_significance>100</concept_significance>
       </concept>
 </ccs2012>

\ccsdesc[500]{Human-centered computing~Information visualization}
\ccsdesc[300]{Mathematics of computing~Graph algorithms}
\ccsdesc[100]{Human-centered computing~Graph drawings}
\ccsdesc[100]{Information systems~Data mining}

\keywords{Co-occurrence patterns, 
Modular Graph Decomposition, 
2-structures, 
Clans, 
Generalized Gaifman graphs, 
Exploratory Data Analysis}

\ENDOMIT

\maketitle

\begin{abstract}
We argue that the existing knowledge about modular
decomposition of graphs and clan decomposition of
2-structures can be put to use advantageously in
a context of data analysis. We show how to obtain
visual descriptions of co-occurrence patterns by
employing these decompositions on possibly
generalized Gaifman graphs associated to datasets.
We provide both theoretical advances that connect 
the proposed process to other data mining aspects 
(namely, closed set mining), as well as implemented 
algorithmics leading to an open-source tool that 
demonstrates our approach.
\end{abstract}

\section{Introduction}

We focus on two known concepts in the Theory of 
Computation, namely, modular decompositions 
and Gaifman graphs, as well as on a known 
generalization of modules, called clans. 
From a perspective of Data Analysis,
modules and Gaifman graphs, taken
together, have the potential of providing 
intuitive hierarchical 
depictions that 
explain data in interesting ways; also, some simple 
generalizations of Gaifman graphs may allow
for clan decompositions and thus enhance 
that potential. 

In order to motivate this work, we report briefly
on one of the cases described in a previous
conference submission~\cite{DBLP:conf/ida/BalcazarPR18}.
We consider a simplified version of the famous 
dataset Titanic\footnote{This variant, prepared by Radford Neal for
the DELVE system, was downloaded from: 
\url{http://www.cs.toronto.edu/~delve/data/titanic/desc.html}
where it is still available at the time of writing.}~\cite{Dawson1995}.
For each of the 2201 people on board the well-known ship,
this dataset records the values of four attributes: 
Traveling Class (first class, second class, third class, 
crew member), Sex (Female, Male), Age (discretized into 
Adult or Child) and Survival (Yes, No), that is, whether 
the person survived the sinking.

\begin{figure}
\centering
\includegraphics[height=0.30\textheight]{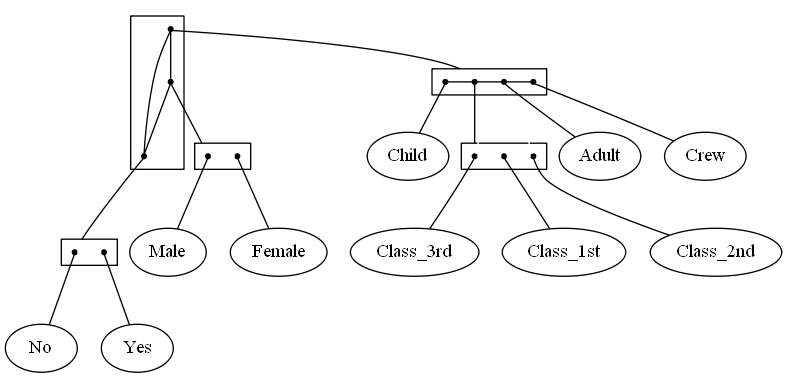}
\caption{Decomposing the standard Gaifman graph of the Titanic dataset.}
\label{fg:titanic}
\end{figure}

The modular decomposition via its standard Gaifman graph 
is shown in Figure~\ref{fg:titanic}. The meaning of each
component of such a figure will be explained in depth along this paper,
as it follows from the definition and properties of modules 
and Gaifman graphs;
but, for the time being, we advance that each black dot represents
a so-called ``strong module''. Some of these are singletons,
which consist of one value of one of the attributes, and the
figure reflects it by linking the black dot to the explicitly
given value. The others are nontrivial modules, represented
by a box linked to the black dot.

Absence of a line connecting two dots means that no value 
below one dot co-occurs ever in the dataset with a value 
below the other: for instance, the two different values of
each of the attributes Survival and Sex cannot occur together,
so they remain disconnected.

On the other hand, a line connecting two dots means that 
every value below one of the dots co-occurs with every 
value below the other, somewhere in the dataset. For instance,
the lower line in the triangle inside the top left box means 
that every combination of values of Sex and Survival appears 
in the dataset. The remainder of the triangle means that each 
of these four values, Yes, No, Male, Female, co-occurs
somewhere with each and every one of the rest of the values: those
of Age and Traveling Class.

The definition of module guarantees that no intermediate 
option exists between full connectivity and no connectivity: 
this is what makes this approach interesting.
Our implemented, open-source system allows users to obtain such representations
out of data with no particular interaction.

We focus now on the rightmost half of the figure.
Like with Sex and Survival, one might expect a ``box'' 
(a module) with the four alternative values of the
Traveling Class attribute, namely, 1st, 2nd, 3rd or Crew. 
Instead, we see that the value Crew ``has migrated'' 
to the parent box that includes the age values.
This calls our attention to the following fact:
whereas both adults and children appear in all the
traveling classes, the crew included, of course, 
no children, at least as public records show;
a fact that we might overlook in a non-systematic 
analysis and that is represented by the absence of
a connecting line between Crew and Child.
That is: even if the traveling class and the ``Crew'' label are
employed as values in the same column, the data tells us,
through our decomposition procedure, that they have different semantics.
Additional discussion can be found in~\cite{DBLP:conf/ida/BalcazarPR18}.

Thus, these hierarchical 
depictions provide often easy-to-grasp
explanations of co-occurrence patterns in the given data. 
Our team has shown also that, in a context of diagnostic data 
from medical practice in a cooperating hospital, the diagrams that we
can obtain from our approach provide visual, interpretable hints
about patterns of diagnostics that co-occur in
the data~\cite{CBMS},~\cite{WomEncourage},~\cite{CISI}. 

The aim of this paper is twofold. We provide here 
a fundamental study of the process behind these 
applications, so as to better understand the 
processes we are applying; and also we develop a 
``systems'' perspective, explaining the key implementation 
ideas employed in our open-source software tool to perform 
this sort of decomposition. Our theoretical study both
supports our algorithmics and, also, explains how our
approach relates to other existing data analysis options.
Specifically, we explain how the modular decompositions 
of standard Gaifman graphs, and also the more general 
clan decompositions of some variants that we describe, 
fit a specific view of closure spaces that correspond 
to so-called ``modular implications'', 
resp.~``clan implications'', that we introduce here. 
This allows us to connect our approach with closure and 
implication mining, which are quite standard tools in 
Data Analysis.

\begin{figure}
\centering
\includegraphics[width=3.5in]{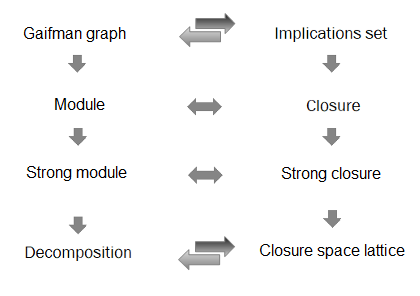}
\caption{General scheme of our approach.}
\label{fg:scheme}
\end{figure}

Figure~\ref{fg:scheme} shows us a general perspective 
of the intuitions behind our development. Both ``vertical
sequences'' of processes are known; on one hand, from any
graph, and thus for the so-called Gaifman graph that one
can associate to a dataset, it is possible to determine 
its modules, filter the so-called strong modules, and
organize them into the modular decomposition of the graph. 
In a similar way, we have that from a set of implications 
we can obtain closed sets and determine which of them are 
strong closures within the corresponding closure lattice:
a well-known process.
 
Throughout this work we develop a theoretical 
framework that connects both processes:
we show how to define
the set of implications associated to a graph in such a
way that strong modules and strong closures are the same 
sets. This connects intimately the modular decomposition
with the closure lattice. As we move on to consider some
extended, more quantitative analysis via generalized 
Gaifman graphs, we shall see that modules become insufficient 
and we adopt a known generalization, namely clan decompositions 
of symmetric 2-structures, for which we develop an analogous 
connection with closure lattices.

We are using different arrows for leftward / rightward 
connections in the top and bottom layers of~Figure~\ref{fg:scheme}.
At the top layer, given any graph, we demonstrate how to
construct the modular implications, and how this process
generalizes to clan implications; however, for the converse,
not every implication set is valid, and it remains open to
find intrinsic characterizations of sets of implications
that correspond to a graph. Moreover, as we discuss below,
the modular decomposition of a graph and of its complement
are isomorphic and, accordingly, the modular implications 
do not distinguish between them.

Similarly, at the bottom layer, the closure space lattice
determines the decomposition but, in the general case, it
includes additional information; that is, the decomposition
may not be sufficient to infer the whole closure space.

\section{Preliminaries}
\label{preliminaries}

Our work builds upon both the known connection between closure
spaces and implications, and on several proposals for identifying
hierarchical decompositions of graphs, namely, modules and, in
a more general way, clans of 2-structures. 
However, the theory of 2-structures
and their clans will be provided later on, after motivating the
need of using them for certain generalizations of Gaifman graphs;
initially, we will work with standard Gaifman graphs and
we will only need modular decompositions. We explain these notions
in this section.

\subsection{Set-Theoretic Notations, Implications, and Closures}

We assume that a given set $\U$ of potential items of interest 
has been fixed. Datasets will be assumed to come in transactional 
form, where each transaction is simply a subset of items. 
Set-theoretic notations are standard; $X\setminus Y$ denotes
set difference: the elements of $X$ that do not appear in $Y$.
The complement of $X$ is~$\U\setminus$$X$. Following customary usage,
union is often denoted by mere juxtaposition, as in $XY$.

\begin{definition}
Two sets $X$ and $Y$ overlap if the sets $X\cap Y$, $X\setminus Y$ and $Y\setminus X$ are all three nonempty.  
Equivalently, $X$ and $Y$ do not overlap if and only if they are either disjoint, or a subset of one another.
\end{definition}


Closure spaces on the powerset of the set of items will play an important role 
in our developments. A closure operator maps each set of items $X$
to a set of items, its ``closure'' $\overline{X}$; 
they are characterized by three properties:
\begin{itemize}
\item Extensivity: $X \subseteq \overline{X}$
\item Idempotency: $\overline{\overline{X}} = \overline{X}$
\item Monotonicity: if $X \subseteq Y$ then $\overline{X} \subseteq \overline{Y}$
\end{itemize}

A set is a \textit{closed set} if it coincides with its closure,
and the closure space is the family of all the closed sets.
A basic fact from the theory of closure spaces is that the
intersection of closed sets is again closed.

Now we let items play the role of propositional variables. Then,
implications are conjunctions of definite Horn clauses, where
a set of clauses sharing the same antecedent are written as
a single formula with conjunctions both at the antecedent and
the consequent of an implication connective; we often write 
antecedents and consequents as mere sets, letting the conjunction
connective implicit. For sets of items $X$, $Y$, and $Z$,
$Z$ satisfies the implication $X\dto Y$, denoted as 
$Z \models X\dto Y$, if either $X \not\subseteq Z$ or $XY \subseteq Z$.

The fact, well-known in logic and knowledge representation, 
that Horn theories are exactly those closed under bitwise 
intersection of propositional models leads to a strong 
connection with closure spaces. It runs as follows: 
given a set of implications $\B$, the closure $\overline{X}$ 
of a set $X$ is the largest set $Y$ such that 
$\B$ 
logically entails $X \Rightarrow Y$; whereas, if we are given a closure 
operator, we can axiomatize it by the set of implications 
$\{X \dto Y: X \subseteq \U, Y \subseteq \overline{X}\}$ or, 
equivalently, any set of implications that entails exactly this set. 

For references and supporting facts of all our claims so far, see the discussions in~\cite{DP},~\cite{KR} and~\cite{Wild} or its early version~\url{https://arxiv.org/abs/1411.6432v2}.

\begin{definition}
\label{df:strongclosure} 
$\overline{X}$ is a strong closure if, for all other closures $\overline{Y}$, $\overline{X}$ and $\overline{Y}$ do not overlap.
That is, either $\overline{X}$ and $\overline{Y}$ are disjoint, or are a subset of one another.
\end{definition}

\subsection{Modular Graph Decompositions}
\label{ssec:modules}

In many cases, items of $\U$ will be vertices of graphs in
a fully standard way. Then, we use the terms ``vertex'' and ``item''
interchangeably.
In all our graphs, we assume that no self-edge is ever present;
also, we do not allow for multiple edges.

The modular decomposition of a graph is a process that 
consists of decomposing the graph into sets of vertices, 
nowadays called ``modules''. It has been rediscovered many times and 
described under many different names~\cite{survey}, but the earliest appearance of
the notion seems to be~\cite{Gallai1967}~\footnote{An 
English translation is available as~\cite{Gallai1967translation}, where they are
called ``homogeneous sets''.}. 

\begin{definition}
Given a graph, a set $X$ of vertices is a module if, 
for each vertex $y \notin X$, 
either every member of $X$ is connected with $y$ 
or every member of $X$ is not connected with $y$. 
\end{definition}

As modules can be proper subsets of other modules, we can obtain   
a hierarchical decomposition of the graph; these facts
have been studied in a very
wide bibliography (see~\cite{survey} and the references there).

The modules of a graph can be seen, therefore,
as subgraphs of the original graph. 
Note that the set of all vertices is vacuously a module, 
as are each vertex by itself and the empty set. These 
are called the trivial modules; we will systematically
ignore the empty module. 

As an intuition device that will become full-fledged in
subsequent sections, we take the somewhat anthropomorphic
stance of considering the presence or absence of an edge
as the way one vertex ``sees another''. Thus, we will
often resort to expressions like 
``an item \textit{sees in a different way} two other items'' 
when it is connected to one of them and disconnected from 
the other one; or, in the same case, we may say that 
``the item \textit{can distinguish} between two other items''. 
This intuition is customary in the field
of 2-structures where the notion of a ``clan'', 
a generalization of modules defined
also below in Section~\ref{sec:2strclans}, relies 
on this intuitive device. 

Hence, throughout this paper, the 
presence or absence of edges will convey only the
idea of two different ways one item sees another; that is,
we give the same interpretation to a graph and to its
complement. Indeed, permuting absence and presence of 
all edges does not change the ways vertices distinguish 
each other. Thus, we may state the following proposition:

\begin{proposition}
\label{prop:graphsamecomplement}
A graph and its complement leave the same set of modules, therefore, the same modular decomposition.
\end{proposition}

On the basis of the definition of module, we can state
(see~\cite{survey}):

\begin{proposition}
\label{prop:basicfacts}
Consider a graph and the modules in it.
\begin{enumerate}
\item
Let $x$ and $y$ be elements of a module $\M$: 
if vertex $z$ is connected to exactly one of $x$ and $y$ 
then, necessarily, $z \in \M$.
\item
Let sets $X$ and $Y$ be non-disjoint: $X\cap Y \neq \emptyset$.
If both are modules, then also $X\cap Y$, $X\setminus Y$, and 
$Y\setminus X$ are modules.
\item
Let sets $X$ and $Y$ be disjoint. If both are modules, then 
either every $x\in X$ is connected to every $y\in Y$, or
no $x\in X$ is connected to any $y\in Y$.
\end{enumerate}
\end{proposition}

This last statement was already announced informally
in our Introduction. It leads to a main interest 
of the notion of module, namely, all the vertices of
a module can be collapsed into a single vertex without
ambiguity with respect to how to connect it to the
rest of the vertices: 
the new vertex gets connected to $y$ if all the
members of the module were connected to $y$, and
remains disconnected if all the members were 
disconnected. 
Clearly, the definition given of module is what 
is needed for this process to be applied without 
ambiguity about whether the new vertex should or 
should not be connected to some external vertex $y$. 
More generally, the same considerations
apply if we simultaneously
collapse into single vertices two disjoint modules: either they
are connected, in the sense that  all the respective pairs 
of vertices (one from each module) are, or they are not 
because no such pair is.

Nothing forbids modules to intersect each other;
in that case, though, collapsing one module into 
a single vertex may affect
the other. In order to avoid side effects, it is 
customary to restrict oneself to so-called 
``strong modules''~\cite{Gallai1967}
(see also \cite{Gallai1967translation}): 
they allow us to obtain a tree-like decomposition. 

\begin{definition}
A module $\M$ is a strong module of a graph if it does
not overlap any other module; that is, for all other modules 
of the graph $\M'$, either $\M \cap \M' = \emptyset$ or they 
are subsets of one another.  
\end{definition}

Given a graph, we can focus on its maximal strong modules; 
it is known that each vertex belongs to exactly one of them~\cite{survey}.
Thus, one, or more, or even all of these maximal strong modules
can be collapsed into a single vertex each. The resulting graph
is called as ``quotient graph''. 

\begin{definition}
\label{df:coarsestquotient} 
Given a graph, its \emph{coarsest quotient graph}
is obtained by collapsing into respective vertices
all its maximal strong modules.
\end{definition}

Then, a tree-like structure arises from the fact that each
of these modules, taken as a set of vertices, is actually 
a subgraph that can be recursively decomposed, in turn, into maximal 
strong modules, thus generating views of subsequent internal 
structures given by their respective coarsest quotient
graphs. We display the decomposition tree while labeling
each node (a strong module) with the corresponding coarsest
quotient graph, and connect visually each collapsed vertex
to the subtree decomposing the corresponding module.
Of course, the root of the ``tree'' is the coarsest
quotient of the whole graph.

\begin{figure}	
\centering
\includegraphics[width=1.5in]{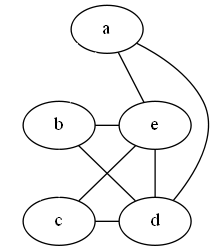}
\centering
\includegraphics[width=1.5in]{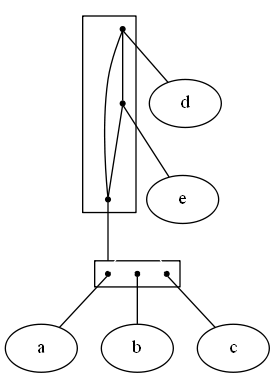}
\caption{A graph and its modular decomposition.}
\label{fg:firstgraph}
\end{figure}

\begin{example}
\label{ex:easy}
Figure~\ref{fg:firstgraph} shows a simple example
of modular decomposition. It is easy to check that
$\{a, b, c\}$ conforms a  module: both $d$ and $e$ 
are connected to each of them. The other possible 
nontrivial modules are $\{a,b,c,d\}$, $\{a,b,c,e\}$, 
$\{d, e\}$, $\{a, b\}$, $\{a, c\}$, and $\{b, c\}$;
they are not strong: each of them intersects others.
Instead, $\{a,b,c\}$ is a strong module as it does not overlap any other module. 
The root is the fully connected
coarsest quotient graph where $\{a, b, c\}$, the single
nontrivial maximal strong module, has been
collapsed to a single vertex. Each of the three vertices
is connected to the module they represent: two are maximal
strong but trivial ones, and the largest one decomposes 
itself again into three trivial modules.
\end{example}

Of course, Figure~\ref{fg:titanic} is also an example of modular
decomposition, to which we will return later on, and 
further examples will keep appearing throughout the paper.

\subsection{Gaifman Graphs}

Gaifman graphs are mathematical structures introduced 
several decades ago as a means to study limitations 
of the expressivity of logical languages \cite{DBLP:books/sp/Libkin04}.
They are defined on first order relational structures (that is, 
in essence, relational databases) and their basic notion is pretty simple: 

\begin{definition}
\label{df:gaifmangraph} 
Given a first order relational structure $\{R_i\}_{i\in I}$ 
where the values appearing in the tuples of the relations $R_i$
come from a fixed universe $\U$, its corresponding Gaifman graph 
has the elements of $\U$ as vertices, and each edge $(x,y)$, 
for $x \neq y$, is present in the graph exactly when $x$ and $y$ appear 
together in some tuple $t \in R_i$, for some $R_i$.
\end{definition}

We extend this notion in various ways along this paper. Labeled versions
will be introduced below in Section 5; for the time being, we only
remark that the notion extends naturally into transactional data:
for each transaction, and for each pair of different items $x$ and 
$y$ in that transaction, we ensure that the edge $(x,y)$ is present
in the graph, and that only such edges are.

It is, of course, possible to apply the modular decomposition method
on this sort of graph. In reference to the data on which the graph was
constructed: what can be said of the obtained decomposition, in terms of
data analysis? This paper belongs to a line of research based on
this question.

\section{A Closure-Based View of Modules} 
\label{sec:closureview} 

We start with the top, left-to-right arrow of Figure~\ref{fg:scheme}:
given a graph, we show how it is possible to describe the conditions for a
subgraph being a module in the form of a set of implications.
This leads to connections between tree-like modular
decompositions, as described above, and the theory of closure spaces.
We show first how to obtain these implications, and then discuss
some of the resulting connections. 

Indeed, the implicational view is already implicit in
Proposition~\ref{prop:basicfacts} (1): if $x$ and $y$ are elements of any module $\M$
and~$z$ sees them in a different way, then $z \in \M$. With the standard
semantics of implications, taking again vertices (items) as propositional variables
and, hence, subgraphs as models, this is equivalent to: $\M \models xy \dto z$.

To pursue this idea further, given a graph, 
we generate a set of implications from it;
we call this set the set of ``modular implications''
of the graph.
In each implication, the antecedent will be a pair of vertices 
in the graph, and the consequent will be conformed by those 
vertices that see the vertices in the left side in different ways.

Formally, we need to establish some notation. 
Given a graph $\G$ and a vertex $x$ in it, 
we denote by $\N_{\G}(x)$ the set of immediate neighbors of $x$ in $\G$,
and define the ``distinguishing set'' of a pair of vertices
$x$, $y$, as follows:
$$
\D_{\G}(x, y) = (\N_{\G}(x) \setminus \N_{\G}(y)) \cup (\N_{\G}(y) \setminus \N_{\G}(x)) \setminus\{x,y\}.
$$
That is, this set gathers together all the vertices
that see one given pair in different ways. Note that it is
necessary to remove explicitly $x$ and $y$ from the 
distinguishing set. Indeed, in case they are connected, 
the absence of self-edges would make each of them qualify 
to distinguish themselves from the other, 
whereas the definition of module only searches for distinguishing 
vertices outside the module.
Then, we construct the set of modular implications as follows:

\begin{definition}
For graph $\G$ and different vertices $x$, $y$ in it, the corresponding
modular implication is
$
xy \dto \D_{\G}(x, y)
$.
The set of modular implications for $\G$ is formed by the modular
implications corresponding to every pair of different vertices,
for which the right-hand side is nonempty: $\D_{\G}(x, y) \neq \emptyset$.
\end{definition}

We can state the following:

\begin{proposition}
\label{prop:easy}
For graph $\G$ and different vertices $x$, $y$ in it, in the corresponding
modular implication
$
xy \dto Z
$
we have $z\in Z$ if and only if $z\notin\{x,y\}$ and $z$ is connected 
to exactly one of $x$, $y$. Therefore, $\{x,y\}$ is a module (of size 2)
if and only if $Z = \emptyset$.
\end{proposition}

The proof is direct from the definition of module. 

\begin{example}
\label{ex:easyagain}
Let us continue the previous example 
by showing the set of modular implications from 
the same graph and the corresponding closure space. From Figure~\ref{fg:firstgraph} we obtain the following set of modular implications:

\begin{multicols}{4}

$ad \dto bc$

$ae \dto bc$

$bd \dto ac$

$be \dto ac$

$cd \dto ab$

$ce \dto ab$

\end{multicols}

The pairs 
$\{ a, b \}$,
$\{ a, c \}$,
$\{ b, c \}$,
$\{ d, e \}$,
are size-2 modules and would lead to 
empty right-hand sides in their modular implications; accordingly,
these implications are discarded.

In the left side of Figure~\ref{fg:closespace} we find the closure space lattice described by the modular implications obtained from the graph. For convenience in the comparison, we display again at the right the decomposition from Figure~\ref{fg:firstgraph}, and we mark in bold those closed sets that do not overlap any other closed set (strong closures
as per Definition~\ref{df:strongclosure}).

\begin{figure}	
\centering
\includegraphics[width=1.75in]{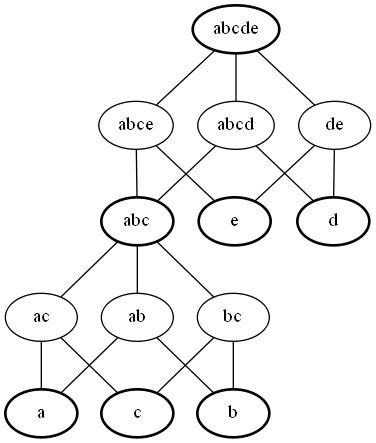}
\centering
\includegraphics[width=1.5in]{graphM2Decomp.png}
\caption{Closure lattice and modular decomposition.}
\label{fg:closespace}
\end{figure}

\end{example}

The main result of this section is Theorem~\ref{th:closuresmodules}.

\begin{theorem}
\label{th:closuresmodules} 
The modules of a graph and the closures defined by its set of modular implications are the same sets. 
\end{theorem}

\begin{proof}
\textit{(Closures are modules.)}
Let $X$ be a closure, and suppose that there is some $y$ that can 
distinguish two arbitrary $x_1,x_2 \in X$; one of the modular implications will be $x_1x_2 \dto Y$
with $y\in Y\neq\emptyset$. 
As $X$ is a closure, it must satisfy all the modular implications, and 
both antecedents are in $X$, thus $y\in X$. Hence, no $y$ outside
$X$ may distinguish two elements inside $X$, which is the definition of module. 

\textit{(Modules are closures.)}
Let $X$ be a module. It suffices to show that 
it satisfies all the modular implications: 
let $x_1x_2 \dto Y$ be one of them. If either $x_1 \notin X$
or $x_2 \notin X$, then $X$ satisfies the implication
by failing the antecedent.
If $x_1, x_2 \in X$ then, since $X$ is a module, 
no item outside $X$ can distinguish them;
but, according to Proposition~\ref{prop:easy}, $Y$ is
the set of items that distinguish them, hence $Y \subseteq X$
and $X$ satisfies the modular implication.
\end{proof}

\begin{corollary}
\label{cr:samestrong}
Strong modules and strong closures are the same sets.
\end{corollary}

Thus, the match seen in Figure~\ref{fg:closespace}
is not chance. The closures that are not strong do
correspond exactly to the additional modules described
at the end of Example~\ref{ex:easy}.
In the following, and until explicitly indicated otherwise,
whenever we talk about closures, we are refering to the 
closures described by the set of modular implications.

\subsection{Graph Reconstruction}

This section develops the reverse arrow at the top of Figure~\ref{fg:scheme}:
from the set of modular implications of a graph $\G$ defined on the vertices 
$\U$$= \{v_1,\ldots, v_n\}$, it is possible to construct a graph $\G_a$ with exactly the same decomposition, taking just those implications whose left parts are in the form $v_1v_2,\ldots,v_{n-1}v_n$.
The graph is either $\G$ or its complement because,
following Proposition\ref{prop:graphsamecomplement},
they have the same set of modules and essentially the same modular decomposition.
The idea to reconstruct the graph is to take from the set of implications a 
smaller number of well-chosen implications; we show that they give us enough information about the relations between the edges to be able to reconstruct the graph.

\begin{algorithm}[t]
\caption{Construction of a connex graph $\G'$ on the set of vertices determined by the edges of a graph $\G$}
\begin{algorithmic}
\STATE \textbf{Input:} Set of modular implications of $\G$.
\STATE \textbf{Output:} $\G' = (\U_{\G'}$,$\E_{\G'})$, with binary labeling on its edges.
\STATE Let $\U_{\G'}$ be an empty set.
\STATE Let $\E_{\G'}$ be an empty set.
\FORALL{$j=\overline{2,n}$}
	\STATE Add $v_1v_j$ to $\U_{\G'}$ 
	\IF {$2<j$}
		\STATE Add $(v_1v_{j-1},v_1v_j)$ to $\E_{\G'}$ in the following way
		\IF {$v_1 \in Z$, such that $v_{j-1}v_{j} \dto Z$}
			\STATE $(v_1v_{j-1},v_1v_j) = 0$
		\ELSE
			\STATE $(v_1v_{j-1},v_1v_j) = 1$
		\ENDIF
		\FORALL{$i =\overline{2,j-1}$}	
			\STATE Add $v_iv_j$ to $\U_{\G'}$ 
			\STATE Add $(v_{i-1}v_j,v_iv_j)$ to $\E_{\G'}$ in the following way
			\IF {$v_j \in Z$, such that $v_{i-1}v_i \dto Z$}
				\STATE $(v_{i-1}v_j,v_iv_j) = 0$
			\ELSE
				\STATE $(v_{i-1}v_j,v_iv_j) = 1$
			\ENDIF
		\ENDFOR
	\ENDIF
\ENDFOR
\end{algorithmic}
\label{alg:form}
\end{algorithm}

Before explaining our algorithm to reconstruct the graph we have the following propositions.

\begin{proposition}
\label{prop:implications}
Let $Z = \D_\G(x,y)$, if we have the module implications $xy\dto Z$  we know $(x,z) \neq (y,z)$ for all $z \in Z$. Also we know, $(x,w)= (y,w)$ for  all $w \in \U \setminus$$Z$
\end{proposition}

The implications will tell us only which vertices distinguish other vertices.
Even if we know that vertex $z$ distinguishes $x$ ftom $y$, this only
tells us that $z$ is connected to exactly one of the other two, but 
we do not know which one. We can choose one of them, but then the choice
influences the presence or absence of other edges.

\begin{algorithm}[t]
\caption{Constructing the path that connects two vertices $v_lv_k$ and $v_mv_p$ in the graph $\G' = (\U_{\G'}$,$\E_{\G'})$}
\begin{algorithmic}
\STATE \textbf{Input:} Two vertices $v_lv_k \in \U_{\G'}$ and $v_mv_p \in \U_{\G'}$, where $k\leq p$, without losing generality.
\STATE \textbf{Output:} The set of vertices that conform the path that connects the vertices $v_lv_k$ and $v_mv_p$ 
\STATE Let $Path$ be an empty set
\IF {$k==p$}
	\IF{$l < m$}
		\FORALL{ $i=l$ to $m$}
			\STATE Add $v_iv_p$ to $Path$
		\ENDFOR
	\ELSE
		\FORALL{ $i=m$ to $l$}
			\STATE Add $v_iv_p$ to $Path$
		\ENDFOR
	\ENDIF
\ELSE
	\FORALL{$i=l$ to $1$}
		\STATE  Add $v_iv_k$ to $Path$
	\ENDFOR
	\FORALL{$i=k+1$ to $p$}
		\STATE Add $v_1v_i$ to $Path$
	\ENDFOR
	\FORALL{$i=2$ to $m$}
		\STATE Add $v_iv_p$ to $Path$
	\ENDFOR
\ENDIF
\STATE Return $Path$
\end{algorithmic}
\label{alg:path}
\end{algorithm}

The way through will be to construct a larger but sparse connex graph $\G'$ using the Algorithm~\ref{alg:form}, whose vertices are all possible edges from the graph $\G$ and the edges determine the relation between these edges:
either one exists in $\G$ and the other does not, or both are equal in this
respect (either by existing both or none of them).
As the graph $\G'$is a connex graph, once a type of edge is assigned 
to any single edge, we may deduce the type for all the remaining edges and construct the graph $\G_a$. We find that the order of the edges in the graph $\G'$ is related to the first two levels of the SE-tree~\cite{DBLP:conf/kr/Rymon92}.

For the Algorithm~\ref{alg:form}, it is possible to determine when two edges get different status by following the Proposition~\ref{prop:implications}: two given edges, $v_iv_k$ and $v_jv_k$ of $\G$, are in different status if for the implication $v_iv_j \dto Z$, $v_k \in Z$.
Whether we know that two edges are in the same status, or in different ones, 
as soon as we can determine whether one of them is an edge of $\G$ 
the other gets determined as well.

The Algorithm~\ref{alg:form} constructs a connex graph since given two vertices on it there is a path that connects them, the Algorithm~\ref{alg:path} gives us this path. In fact, the length of the path that connects the edge $v_lv_k$ to $v_mv_p$ is $(l-1)+(p-k)+(m-1)$ where $p > k$, while if $p=k$, the length of the path is $|l-m|$. 

\begin{example}
\label{ex:reconstructgraph}

Let us apply the Algorithm~\ref{alg:form} on the following set of implications obtained from the graph of Figure~\ref{fg:reconstructgraph}:

\begin{multicols}{4}

$ab \dto cd$

$ac\dto be$

$ad \dto b$

$ae \dto bd$

$bc \dto de$

$bd \dto c$

$be \dto ac$

$cd \dto e$

$ce \dto ad$

$de \dto a$

\end{multicols}

Thus, Algorithm~\ref{alg:form} leads to 
the connex graph $\G' = (\U_{\G'}$,$\E_{\G'})$, 
where $\U_{\G'}$$=\{ab,ac,bc,ad,bd,cd,ae,be,ce,de\}$,
as displayed in Figure~\ref{fg:simil}. A label 1 joining
two vertices of $\G'$ means that both of the 
corresponding edges of $\G$ must be in equal status:
either both are present or both are absent in $\G$.
A label 0, instead, means that they must have 
different status: exactly one of them exists in $\G$.

\begin{figure}[t]
\centering
\includegraphics[height=0.2\textheight]{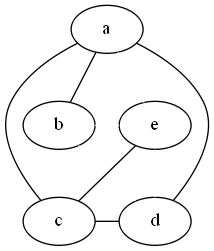}
\caption{Initial graph for Example~\ref{ex:reconstructgraph}}
\label{fg:reconstructgraph}
\end{figure}

\begin{figure}[t]
\centering
\includegraphics[height=0.35\textheight]{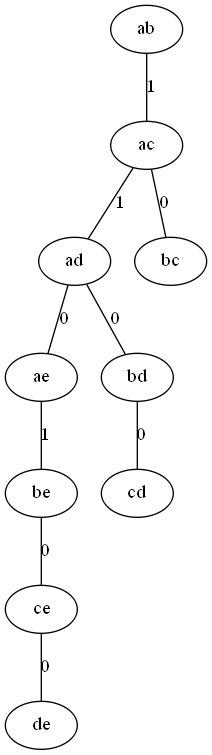}
\caption{Resulting graph for the Algorithm~\ref{alg:form} for the graph on Figure~\ref{fg:reconstructgraph}}
\label{fg:simil}
\end{figure}

The possible reconstructed graphs are shown in Figure~\ref{fg:reconstructedgraph}.
It remains to decide on one edge, such as the initial vertex 
of $\G'$ corresponding to $(a,b)$.

So, suppose the edge $(a,b)$ exists, that is $a$ and $b$ are connected. As we may see in Figure~\ref{fg:simil} $(a,c)$ and $(a,d)$ are in the same status as $(a,b)$, while $(b,c)$ is not, thus $b$ and $c$ are disconnected, and $a$ is connected with $c$ and $d$.
We also know that $(a,d)$ is in different status than $(a,e)$ and $(b,d)$, thus $a$ and $e$ are disconnected, and also, $b$ and $d$ are disconnected.
We know $(b,d)$ and $(c,d)$ are in different status, thus, $c$ and $d$ are connected;
while $(a,e)$ is in the same status as $(b,e)$, thus, $b$ and $e$ are disconnected.
Also, we have that $(b,e)$ is different from $(c,e)$, meaning that $c$ and $e$ are connected. At the same time $(c,e)$ is different from $(d,e)$, thus $d$ and $e$ are disconnected.
In this way we get the graph in the left of Figure~\ref{fg:reconstructedgraph}.

If we had started by choosing that the edge $(a,b)$ does not exist, following the constraints describe in Figure~\ref{fg:simil}, we get the graph in the right of Figure~\ref{fg:reconstructedgraph}
which is, of course, the complement graph and has the same set of 
modular implications by Proposition~\ref{prop:graphsamecomplement}.

\begin{figure} 
\centering
\includegraphics[height=0.2\textheight]{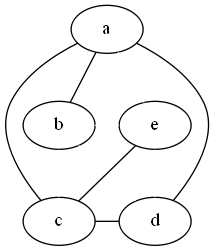}
\centering
\includegraphics[height=0.2\textheight]{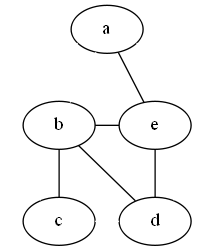}
\caption{Reconstructed graph}
\label{fg:reconstructedgraph}
\end{figure}

\end{example} 

\subsection{Module Taxonomy}

To maintain consistency within the paper, for some notions
we will be using terminology 
corresponding to the theory of 2-structures
\cite{DBLP:books/daglib/0025562}
instead of the classical terms in modular decompositions.
A graph of three or more vertices is called \emph{primitive} if all of its modules are trivial, while if either all of its vertices are connected, or all of its vertices are disconnected, we say that it is \emph{complete}; 
in this case, every subset is a module and there are no strong nontrivial modules.
By convention, graphs of size 1 or 2 are considered complete.
In fact, it can be seen easily by exhaustive inspection
that at least 4 vertices are necessary for primitive graphs.

Modules can be primitive or complete themselves as well.
The qualifier is customarily chosen 
according to their corresponding coarsest quotient graphs
(Definition~\ref{df:coarsestquotient}), and not according 
to themselves as subgraphs. 
Indeed, these terms are most often applied to coarsest quotient graphs,
such as those labeling the nodes of a tree decomposition.
Specifically, it is a theorem of modular decomposition that
the coarsest quotient graphs labeling each node of a
decomposition tree, if nontrivial, are all either primitive or complete.
Graphs whose tree does not show any primitive quotient graph
are called ``cographs'' and have a large number of graph-theoretic
and algorithmic properties.
The graph
$P_4$ is the path on 4 vertices with 3 edges; its complement is also $P_4$,
and plays a key role in modular decompositions, as follows:

\begin{proposition}
\label{prop:P4}
Primitive graphs (and thus
also their complements) always have
induced $P_4$ subgraphs~\cite{DBLP:journals/dam/Spinrad92}\footnote{See
also \url{https://en.wikipedia.org/wiki/Cograph}
and the references there.}.
Hence, for a primitive module, its corresponding coarsest quotient graph consists of at least four maximal strong modules.
\end{proposition}


As an example, both nonsingleton quotient graphs in Figure~\ref{fg:firstgraph} are complete, even if the graph itself is not, whereas in Figure~\ref{fg:titanic}
we can see a primitive coarsest quotient that coincides, actually,
with $P_4$. We show below one example where only primitive modules
appear in the decomposition.

Among other terms, primitive modules are sometimes called
neighborhood modules. Modular decomposition theory distinguishes
fully disconnected complete graphs (or ``parallel'' modules)
and fully connected ones (or ``series'' modules), leading to
often duplicated arguments and definitions because both cases
fulfill the same role. Indeed, recall from Proposition~\ref{prop:graphsamecomplement} 
that presence or absence of edges can be swapped with no change in the modular structure: that's why fully connected and fully disconnected modules are treated similarly. Hence, we prefer the  
view, originated in the study of 2-structures, of naming them both ``complete''
\cite{DBLP:books/daglib/0025562} (see below).

\subsection{Closures and Module Types}

By Theorem~\ref{th:closuresmodules} and Corollary~\ref{cr:samestrong} we have that modules and closures, strong or not, are the same sets, this relation is represented in the Figure~\ref{fg:scheme} by the double arrows.
As we shall see below, the strong modules and their types will be
important in the visualization of co-occurrence patterns in data
analysis. Thus,
we would like to be able to get the type of a strong module, that is, whether the module is complete or primitive, by using just the information in the closure lattice. 

Following this idea we show, in this section, that the type of a strong module is given by the immediate closure subsets of its respective strong closure; in this way, we may go from the modular decomposition of a graph to the closure space lattice, and viceversa, as the arrows at the end of our scheme in the Figure~\ref{fg:scheme} indicate.

The immediate closure subsets of a set $X$ are the children of $X$ in the closure lattice: namely, closed (not necessarily strong) subsets $Y$ of $X$ such that no intermediate set $Z$, $Y\subset Z\subset X$, is closed.

\begin{theorem} 
\label{th:moduletype}
The type of the coarsest quotient graph of a module is primitive if and only if the immediate closure subsets of its corresponding closure in the closure lattice are strong closures and they are more than three.
\end{theorem}

\begin{proof}
Let $\M$ be a module; by Corollary~\ref{cr:samestrong} its corresponding closure in the closure lattice is also $\M$,
and let us suppose that its coarsest quotient graph collapses the maximal strong modules $\M_i$
($i\in I$ for some index set $I$).

($\Rightarrow$)
Let $\M$ be a primitive module, that is its corresponding coarsest quotient graph is primitive; we have to prove that there is not any union of $\M_i$'s closure subset $\F$ such that $\M_i \subset \F \subset \M$. 
$\F$ must be a union of $\M_i$ since, by the premise, $\M_i$ are strong closures so $\F$ does not overlap any of them.
 
But if $\F$ includes at least two $\M_i$, by definition of primitive there must be at least one other $\M_i$ 
distinguishing them; thus $\F$ could not be a closure. 

($\Leftarrow$)
We have to prove that if the immediate closures $\M_i$ of a strong closure $\M$ are strong closures and more than three, then $\M$ is a primitive module. Let $J\subset I$ be the
index set of a proper subset of the modules $\M_i$, consisting of at least two of them
(we can guarantee this possibility by the premise). 
Suppose that all the remaining $\M_i$'s, $i\notin J$, cannot distinguish between them, 
so $\bigcup_{j\in J} \M_j$ must be a closure;
this would contradict the fact that all the $\M_i$'s are immediate closed subsets of $\M$. Thus, for arbitrary $J\subset I$, there must be at least one $\M_i$, $i\notin J$, that may distinguish between some of them so that no such $\bigcup_{j\in J} \M_j$ is a module.
Subsets that are not in the form $\bigcup_{j\in J} \M_j$ are not modules either because
each $M_i$ being a strong module means that no module overlaps any of them.
Thus, $\M$ is a primitive module. 
\end{proof}

Let us see some examples to illustrate both cases. 

\begin{example}

\begin{figure}
\centering
\includegraphics[width=1.25in]{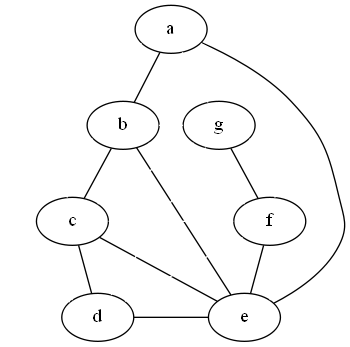}
\centering
\includegraphics[width=1.5in]{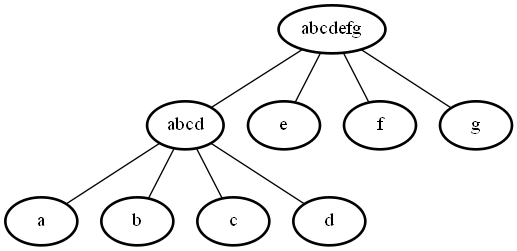}
\centering
\includegraphics[width=1.6in]{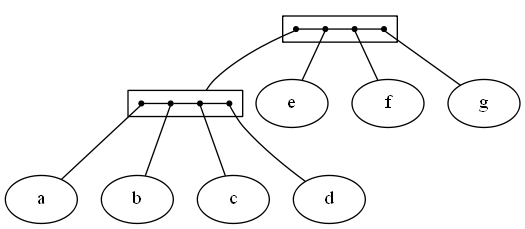}
\caption{A graph with primitive modules, namely $P_4$, into its decomposition tree.}
\label{fg:moduleprimitive}
\end{figure}

In Figure~\ref{fg:moduleprimitive} (left) there is a graph 
with primitive quotient graphs in its decomposition. 
We get from it the following set of modular implications:

\begin{multicols}{4}

$ab \dto c$

$ac \dto d$

$ad \dto bc$

$ae \dto cdf$

$af \dto bg$

$ag \dto bef$

$bc \dto ad$

$bd \dto a$

$be \dto df$

$bf \dto acg$

$bg \dto acef$

$cd \dto b$

$ce \dto af$
 
$cf \dto bdg$

$cg \dto bdef$

$de \dto abf$ 

$df \dto cg$

$dg \dto cef$

$ef \dto abcdg$

$eg \dto abcd$

$fg \dto e$

\end{multicols}

This implication set generates the closure lattice in the center of Figure~\ref{fg:moduleprimitive}. The nodes $abcdefg$ and $abcd$ have four strong closures as children, and indeed their equivalent strong modules into the decomposition are primitive as Theorem~\ref{th:moduletype} says.   
\end{example}

\begin{example}
\begin{figure}[h]
\centering
\includegraphics[width=1.1in]{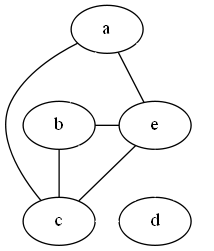}
\centering
\includegraphics[width=1.2in]{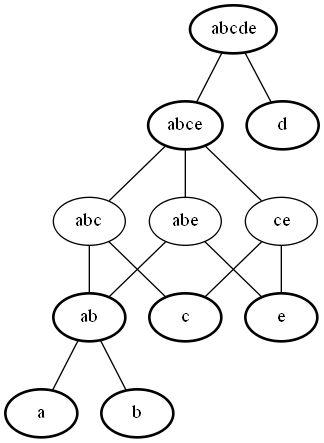}
\centering
\includegraphics[width=0.7in]{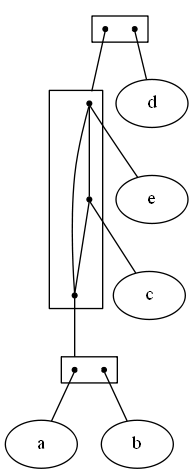}
\caption{A graph with complete modules into its decomposition tree.}
\label{fg:modulecomplete}
\end{figure}

In Figure~\ref{fg:modulecomplete} there is an example of a graph with only complete quotient graphs in its decomposition.
From Figure~\ref{fg:modulecomplete} left we get the set of modular implications:

\begin{multicols}{4}
$ac \dto b$

$ad \dto ce$

$ae \dto b$

$bc \dto a$

$bd \dto ce$

$be \dto a$

$cd \dto abe$

$de \dto abc$

\end{multicols}
with $ab$ and $ce$ closed sets. 

At the center of Figure~\ref{fg:modulecomplete} we have the closure lattice from this set of implications. As we can see, the closed set $abcde$ has two children so its respective quotient graph in the decomposition is a complete module. The closed set $abce$ has three children overlapping, by the Theorem~\ref{th:moduletype}, we may deduce its respective coarsest quotient graph in the decomposition is complete as we can see in the Figure~\ref{fg:modulecomplete} right. The last closed set in the closure lattice is $ab$, it has two children so its respective quotient graph is a complete module as we may see in the decomposition tree.  
\end{example}

\subsection{Data Analysis via Gaifman Graph Decomposition} 

We will return below to a more general approach for these
sort of decompositions. However, we find it is
high time to pause the theoretical development and
motivate, with some example tasks of data analysis, what
we have presented so far in more detail than the mere hints
given in the Introduction. For a given relational dataset, 
we take all values of the attributes (possibly including
an indication of the intended attribute) as items into
a transactional view and, hence, as vertices of a
Gaifman graph. Then, by definition, it represents 
information about item co-occurrences in the dataset,
since edges $(x,y)$, for $x \neq y$, exist when the 
items $x$ and $y$ appear together in some tuple. 
Despite its multirelational potential, to be explored 
in future work, we restrict ourselves to Gaifman graphs 
of single tables, which is sufficient for the commonly 
employed benchmark datasets.
Then, the modular decomposition groups the vertices in 
the hierarchical way that we have seen.

Let us start by going back to the modular decomposition of the standard Gaifman graph of the Titanic dataset that we have described in the Introduction (Figure~\ref{fg:titanic}). The modules for sex and survival are clear and intuitive: as they are different possible values for the same attribute, they never appear together, but happen to have the same set of  neighbors\footnote{Vertices with the same set of neighbors are commonly called ``twins'' in graph theory:
``true twins'' if they are connected and ``false twins''
otherwise, like here; we will see often such cases in our examples
because the same set of neighbors implies that they form a size-2 module.}.
We did already explain why the top module is fully connected: 
a complete coarsest quotient of the whole graph; we also pointed out
how the lack of children among the crew results in the vertex for
Crew moving away from the other values of the same attribute
and into the ``age" module, thus showing the semantic difference
with the traveling classes. The coarsest quotient of this  
module collapses the traveling classes into a single vertex and, 
together with the other three trivial clans, creates
a size-4 primitive module that, by Proposition~\ref{prop:P4}, cannot be
anything else but $P_4$.
Further related discussions are given in \cite{DBLP:conf/ida/BalcazarPR18}.

\begin{figure}[h]
\centering
\includegraphics[width=2.5in]{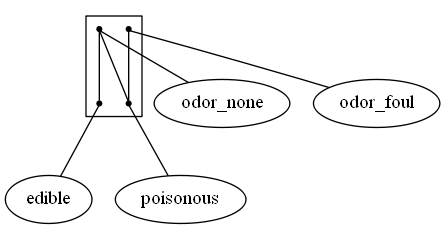}
\caption{A primitive module in the Mushroom dataset decomposition.}
\label{fg:mushroom1}
\end{figure}

Now, we describe some outcomes of analysis on the
well-known dataset Mushroom, also known as ``Agaricus - 
Lepiota''. There, a number of purported attributes of
potential mushrooms of these families are expected to be
useful to predict whether each of the 8124 observations 
would correspond to an edible or a poisonous mushroom~\cite{Dua:2019}.
(The data was compiled by mushroom experts out of hypothetical 
mushroom observations, not actual ones.)

Often, in practice, graphs may have large amounts 
of vertices. Here and also later, with other datasets,
we choose to work with those vertices that appear in the 
transactions more frequently than some determined threshold.
In order to get understandable, smallish 
but representative diagrams; 
we restrict ourselves to the $n$ most frequent
items, for a reasonable value of $n$ that leads to
understandable diagrams. Specifically,
for Mushroom, 
we restrict ourselves first to items appearing at least 2000
times. 

Even then, we discuss some of its modules but
do not display the complete decomposition tree,
because the space is not enough to show it adequately; for
instance, we obtained as a root of the tree decomposition a complete node with many singleton items. Also, of course, a number of false
twins appear, such as grass-living versus wood-living mushrooms,
obviously disconnected. Then, there are also a few interesting 
modules. First, again a $P_4$ case is shown in Figure~\ref{fg:mushroom1}:
if a mushroom has foul odor, then it is a poisonous mushroom,
as ``foul odor'' never appears in the same tuple with ``edible''
(and again the three missing edges are also a $P_4$).

\begin{figure}	
\centering
\includegraphics[width=4.5in]{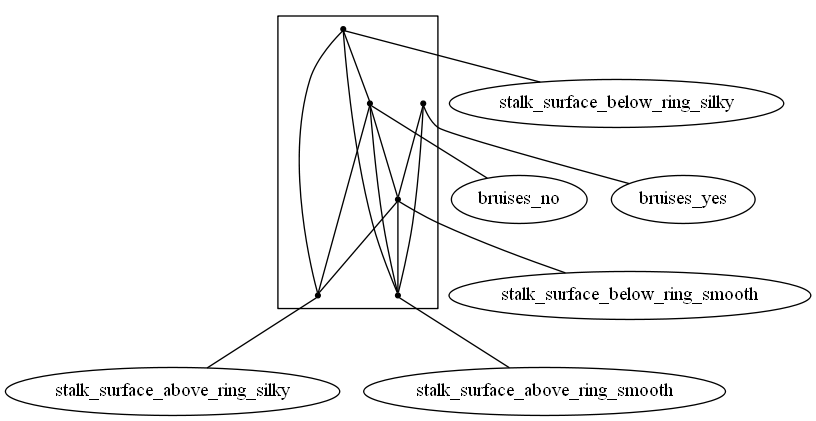}
\caption{A second non-trivial module in the Mushroom dataset decomposition.}
\label{fg:mushroom2}
\end{figure}

Figure~\ref{fg:mushroom2} is a little bit more difficult to 
interpret but we find there, for example, that bruised mushrooms have smooth stalk surfaces, both below and above, but 
never exhibit silky surfaces either below or above, and not all smooth stalk surfaces, either below or above, are bruised, since both
are also connected to the no-bruises vertex.
The induced $P_4$ 
predicted by the theory is indeed there as well (but somewhat more difficult
to pinpoint).

The reader can find further modular decomposition of other datasets 
in our previous papers~\cite{DBLP:conf/ida/BalcazarPR18} and~\cite{CBMS}.

\subsection{Decomposing Thresholded Gaifman Graphs}

Now we consider an alternative way to handle datasets via a variant
of Gaifman graph. 
Sometimes, we may be interested in keeping track of quantitative information that the standard Gaifman graph lacks.
Perhaps the simplest alternative variant of Gaifman graph for 
this purpose is the thresholded Gaifman graph~\cite{DBLP:conf/ida/BalcazarPR18} 
(other alternatives are 
discussed below in Section~\ref{ssec:ggg}).
In this variant, the difference in the edges, instead of being zero joint occurrences versus 1 or more, resorts to a threshold possibly different from 1:
those pairs of items that appear together less often than the determined threshold are considered as ``not frequent enough'' and remain disconnected, whereas connections represent frequencies of co-occurrence higher than or equal to the threshold.

Often, we will apply an additional visual simplification.
With larger datasets, the usual outcome of the analysis
includes mostly large primitive modules of unclear
interpretability. The diagram 
may become, then,
uninteresting. In order to construct helpful visualizations,
we encompass uninteresting substructures into 
single nodes that we label ``Others'', as we will do next.
Two very common cases are, first, large sets of fully disconnected 
vertices and, second, large, complicated primitive modules
that do not convey intuitive information.

\begin{figure}[t]
\centering
\includegraphics[width=1.4in]{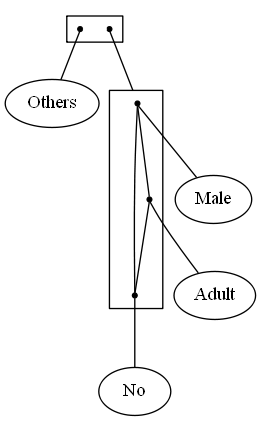}
\caption{Titanic thresholded Gaifman graph decomposition, at 1000.}
\label{fg:titanicthresholded}
\end{figure}

A first example from~\cite{DBLP:conf/ida/BalcazarPR18}:
for the same version of the Titanic dataset described above, assume that we wish to ignore co-occurrences of pairs of items if they occur together less than 1000 times. In the thresholded Gaifman graph, such pairs of items will be kept disconnected. The resulting decomposition tree for that Gaifman graph (slightly simplified by grouping as Others the rest of the fully disconnected vertices as just indicated) 
is shown in Figure~\ref{fg:titanicthresholded}. In a sense, this 
image displays the well-known saying: ``women and children first".

\begin{figure}[t]
\centering
\includegraphics[width=1.8in]{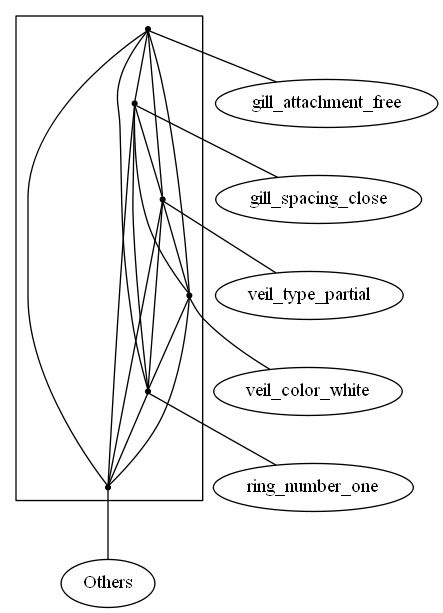}
\caption{Mushroom thresholded Gaifman graph decomposition, at 1000.}
\label{fg:mushroomthresholded1000}
\end{figure}

\begin{figure}[t]
\centering
\includegraphics[width=4.25in]{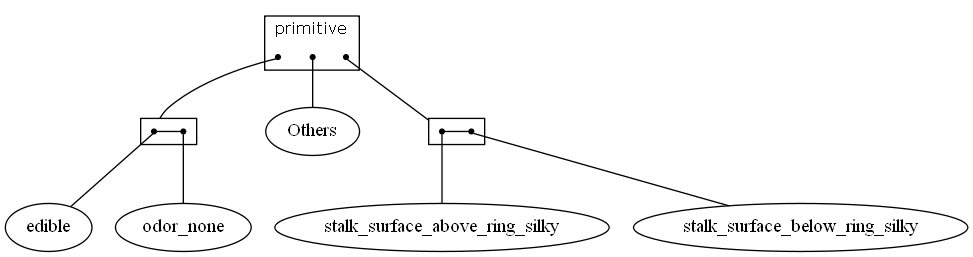}
\caption{Modified module applying the threshold at 800 on the Mushroom dataset.}
\label{fg:mushroomthresholded800cu}
\end{figure}

Working instead with the Mushroom dataset, with attributes appearing more than 2000 times and considering as not frequent enough those co-occurrences below 1000 times we obtain the decomposition shown in Figure~\ref{fg:mushroomthresholded1000}, where we see a handful of
very frequent items that co-occur with each other. Naturally, we might like to see more detailed information: in this case, 
one way to decompose the node Others is to give a lower threshold. Giving 800 as threshold value we may see part of the internal behavior of the Others node; the Figure~\ref{fg:mushroomthresholded800cu} show this part of the decomposition. For this example, we have that there are more than 800 mushrooms but less than 1000 that are edible and do not have odor, and also there is a similar quantity of mushrooms whose ring stalk surfaces, above and below, are silky.

\section{2-structures and clan decomposition}
\label{sec:2strclans}

The notion of modular decomposition is enough to be applied on standard 
or thresholded
Gaifman graphs, but is insufficient to handle adequately other variations of Gaifman graphs that we will want to study as well; for the rest of this paper, we will work with a more general notion, namely, 2-structures and their clans~\cite{DBLP:books/daglib/0025562}.

\begin{definition}
A \textit{2-structure} is a complete graph, endowed with an equivalence relation among its edges.
\end{definition}

In specific, we work only with the so called 
symmetric 2-structures~\cite{DBLP:books/daglib/0025562} 
where all the equivalence classes are symmetric. An equivalence class $\E_i$ on a set $\U$ is symmetric if for all $x,y \in \U$: $(x,y) \in \E_i  \Leftrightarrow (y,x) \in \E_i$. 
Since in the 2-structures that we use the edges represent co-occurrences, is the same to say that $y$ co-occurs with $x$ than $x$ co-occurs with $y$. Then, our equivalence classes of co-occurrences will always be symmetric equivalence classes.
Hence, all along this paper, whenever we talk about 2-structures, 
we will be referring to symmetric 2-structures.

It is possible to see a Gaifman graph as a 2-structure where all the absent edges will be in the same equivalence class while the existing edges will belong to the other equivalence class. In this way we get a complete graph with two equivalence classes. 
In a similar way, thresholded Gaifman graphs also can be seen as 2-structures with two equivalence classes.
Actually, the general form of 2-structures would be akin to directed graphs and,
as Gaifman graphs and their generalizations are instead undirected,
symmetric 2-structures suffice.

This view is also consistent with the claim that in the field of
modular decompositions, the crucial notion is not
presence or absence of an edge, but the difference between them. 
In this sense, a graph and its complement correspond, in fact, 
to the same 2-structure.
The advantage of working with 2-structures is that they allow us to work with more than two equivalence classes. Then, the decomposition procedure is based,
instead of modules, on the so-called ``clans''.
They extend the intuitive concept of when a vertex \textit{sees in different ways} two vertices. In 2-structures we say that an item $x$ ``sees in a different way'' two other items if the edges that connect $x$ with these two items are not equivalent. That is, $x$ sees in a different way $y$ and $z$ if the edges $(x,y)$ and $(x,z)$ lie in different equivalence classes. Alternatively, we say $x$ can distinguish between $y$ and $z$.
Note that this is consistent with the previous usage when a graph is seen as a 2-structure with two equivalence classes.

Thus the notion now corresponding to ``module'' has been called always a ``clan''.
For a 2-structure given by a set of vertices $\U$ and an equivalence relation $\E$ on its edges, we say that the subset $X \subseteq \U$ is a clan, informally, if, 
for every $y \in \U$$\setminus X$, $y$ cannot distinguish the elements of $X$. Formally~\cite{DBLP:books/daglib/0025562}:

\begin{definition}
Given $\U$ and an equivalence relation $\E\subseteq ((\U\times \U) \times (\U\times \U))$ on the edges of the complete graph on $\U$, $C\subseteq \U$ is a clan when 
$$
\forall x\notin C \,\forall y \in C \,\forall z \in C \,((x,y),(x,z))\in \E.
$$
\end{definition}  

Thus, two members of a clan cannot be distinguished by anyone outside the clan,
an idea that motivated the choice of the term.
As with the trivial modules, the so called trivial clans are all the singletons $\{x\}$ for $x\in \U$, as well as $\U$ itself and the empty set.
When we see a graph as a 2-structure, as indicated above, its clans are the modules.
Of course the clans can be again of two types, complete or primitive, like the modules: in a
primitive clan, there are no nontrivial clans; in a complete clan, all edges are equivalent. 
In order to  display the decomposition into a tree-like form we look again for non-overlapping clans.

\begin{definition}
For a fixed 2-structure on universe $\U$, a clan $X \subseteq \U$ is a strong clan if 
$X$ does not overlap any other clan.
\end{definition}

Originally, strong clans were called ``prime clans''. 
However, in the context of modular decompositions, the adjective
``prime'' has received other usages in the literature. We have
deemed better to avoid that adjective, so that previous exposure
of the reader to either modular decompositions or 2-structures
does not result in misunderstandings.

As with modules, strong clans can be collapsed into single
vertices without any ambiguity about how the 2-structure looks like
after the collapse: from the perspective of an outside vertex,
or indeed of a disjoint second clan, all the edges connecting
to nodes inside the clan are of the same equivalence class,
so that class can be chosen for edges upon collapsing
disjoint clans into vertices. The corresponding notion of coarsest quotient
2-structure follows by the same procedure as with modules:
each maximal strong clan collapses to a single vertex.
Then, each of these clans is decomposed recursively so as
to obtain a tree-like decomposition much like those we have
already seen with modules.
We say that $X\subset Y$ is a subclan of $Y$, or alternatively $Y$ is a superior clan of $X$, if $X$ is a maximal strong clan and thus belongs to the coarsest quotient of $Y$.

The internal structure of a clan is also a 2-structure, 
the sub-2-structure that involves the nodes into the clan,
most often organized according to the corresponding 
coarsest quotient.
According to the internal 2-structure of its coarsest quotient, 
each clan can be classified as a complete or a primitive clan:
in a complete clan, all the edges are in the same equivalence class, 
every subset is a clan, and there are no strong clans;
while in a primitive clan there are no nontrivial clans.
It is a theorem of the theory of 2-structures that the nodes of the 
clan decomposition of a symmetric 2-structure 
are all primitive or complete clans. This is the natural
generalization of the modular decomposition of graphs.
Recall that this paper only employs symmetric 2-structures all along;
in the general case, a third type of clan may appear in the
decomposition, namely linear clans.

\section{Closures from 2-structures}
\label{sec:2}

In the same way as we did with modules, it is possible to obtain the set of implications that describe a 2-structure. Again, from the very definition of a clan, we can state: if $x$ and $y$ are elements of any clan $\C$ and $z$ sees them in a different way, then $z \in \C$. This is equivalent to $\C \models ab \dto c$, taking again the standard semantic of implications, the vertices in the graph as propositional variables, and subgraphs as models. We will call this kind of implications ``clan implications''.
 
\begin{figure}
\centering
\includegraphics[width=1in]{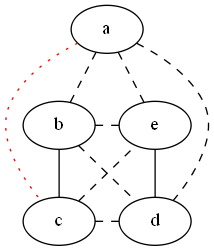}
\caption{A 2-structure.}
\label{fg:2struct}
\end{figure}

Let us see an example of how to get the sets of implications from a 2-structure and show its corresponding closure space. In an illustrative way, it is customary 
to use different colors for each equivalence class, 
so as to enliven the display of 2-structures; to this, we add
a representation of the different equivalence classes by different 
types of lines, taking into account possible b/w printed versions of this paper.
Actually, we often refer informally to the equivalence class of an edge
as the ``color'' of the edge.

From the Figure~\ref{fg:2struct} we obtain the following clan implications set:  

\begin{multicols}{3}

$ab \dto c$

$ac \dto b$
 
$ad \dto ce$

$ae \dto cd$

$bc \dto a$

$bd \dto ce$

$be \dto cd$

$cd \dto abe$

$ce \dto abd$

\end{multicols}

\begin{figure}[h]
\centering
\includegraphics[width=1.8in]{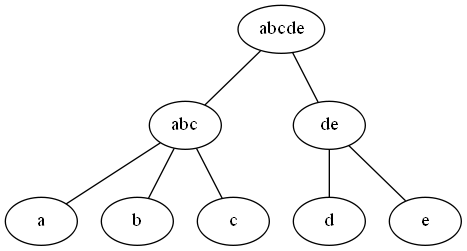}
\centering
\includegraphics[width=1.3in]{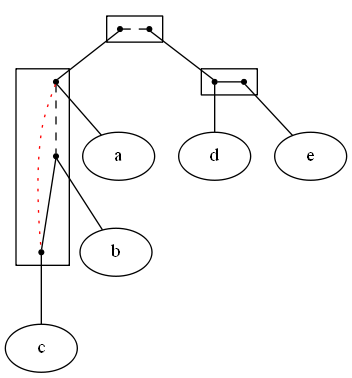}
\centering
\caption{Closure lattice and clan decomposition of the 2-structure in Figure~\ref{fg:2struct}.}
\label{fg:closespace2str}
\end{figure}

The pair $de$ is a closed set.
In Figure~\ref{fg:closespace2str} we find the closure space lattice (left) described by the clan implications set and the clan decomposition tree (right) of the graph in Figure~\ref{fg:2struct}.  

Formally, given a 2-structure $\G$ and a vertex $x$ in it, 
we denote by $\E_{i,\G}(x)$ the set of vertices connected with $x$ by 
edges belonging to
the equivalence class $\E_{i}$ in $\G$,
and define the ``distinguishing set'' of a pair of vertices
$x$, $y$, as follows:
$$
\D_{\G}(x, y) = \bigcup((\E_{i,\G}(x) \setminus \E_{i,\G}(y)) \cup (\E_{i,\G}(y) \setminus \E_{i,\G}(x))) \setminus\{x,y\} 
$$

This set collects together all the vertices that see one given pair in different ways. Then, we construct the set of clan implications as follows:

\begin{definition}
For a 2-structure $\G$ and different vertices $x$, $y$ in it, the corresponding
clan implication is
$
xy \dto \D_{\G}(x, y)
$.
The set of clan implications for $\G$ is formed by the clan
implications corresponding to every pair of different vertices,
for which the right-hand side is nonempty: $\D_{\G}(x, y) \neq \emptyset$.
\end{definition}

From the definition of clan we can state the following:

\begin{proposition}
\label{prop:easyclan}
For a 2-structure $\G$ and different vertices $x$, $y$ in it, in the corresponding clan implication
$
xy \dto Z
$
we have $z\in Z$ if and only if $z\notin\{x,y\}$ and $z$ is connected to each 
of $x$, $y$ by inequivalent edges. 
Therefore, $\{x,y\}$ is a clan (of size 2) if and only if $Z = \emptyset$.
\end{proposition}

\subsection{Connection with Closure Spaces}

We may extend Theorem~\ref{th:closuresmodules} and Theorem~\ref{th:moduletype} to clans in a natural way, getting the theorems:

\begin{theorem}
\label{th:closuresclansclantype}
\begin{enumerate} 
\item Closures from the clan implications are exactly clans, implying as well that strong clans and strong closures are the same sets.
\item The type of a clan is primitive if and only if its immediate subset clans are strong clans and they are more than two.
\end{enumerate}
\end{theorem}

\begin{proof}
1. \textit{(Closures are clans.)}
Let $X$ be a closure, and suppose that there is some $y$ that can 
distinguish two arbitrary $x_1,x_2 \in X$; one of the clan implications will be $x_1x_2 \dto Y$
with $y\in Y\neq\emptyset$. 
As $X$ is a closure, it must satisfy all the clan implications, and 
both antecedents are in $X$, thus $y\in X$. Hence, no $y$ outside
$X$ may distinguish two elements inside $X$, which is the definition of clan. 

1. \textit{(Clans are closures.)}
Let $X$ be a clan. It suffices to show that 
it satisfies all the clan implications: 
let $x_1x_2 \dto Y$ be one of them. If either $x_1 \notin X$
or $x_2 \notin X$, then $X$ satisfies the implication
by failing the antecedent.
If $x_1, x_2 \in X$ then, since $X$ is a clan, 
no item outside $X$ can distinguish them;
but, according to Proposition~\ref{prop:easyclan}, $Y$ is
the set of items that distinguish them, hence $Y \subseteq X$
and $X$ satisfies the clan implication.

2. ($\Rightarrow$) Let $\C$ be a clan by the first part of this theorem, we may state that its corresponding closure in the closure lattice is also $\C$,
let us suppose its coarsest quotient graph collapses the maximal strong clans $\C_i$.
Let $\C$ be a primitive clan, that is its corresponding coarsest quotient graph is primitive; we have to prove that there is not any closed subset $\F$ such that $\C_i \subset \F \subset \C$.
Assume there is such an $\F$.

$\F$ must be a union of $\C_i$'s since, by the premise, $\C_i$ are strong closures so $\F$ does not overlap any of them. Also, 
$\F$ includes at least two of them. By definition of primitive, there must be at least one other $\C_i$ distinguishing them; 
we can pick $y\notin\F$ in the distinguishing clan and $x_1$ and $x_2$ in the distinguished ones inside $\F$: then $\F$ fails
the clan implication $x_1x_2 \dto Y$ and is not a closure. 

2. ($\Leftarrow$)
We have to prove that if the immediate closures $\C_i$, with $i\in I$ 
for some index set $I$, of a strong closure $\C$ are strong closures 
and more than three, then $\C$ is a primitive clan. Let $J\subset I$ 
be the index set of a proper subset of the clans $\C_i$, consisting of at least two of them 
(we can guarantee this possibility by the premise).

Suppose that all the remaining $\C_i$'s, $i\notin J$, cannot distinguish between them, 
so $\bigcup_{j\in J} \C_j$ must be a closure;
this would contradict the fact that all the $\C_i$'s are immediate closed subsets of $\C$. Thus, for arbitrary $J\subset I$, there must be at least one $\C_i$, $i\notin J$, that may distinguish between some of them so that no such $\bigcup_{j\in J} \C_j$ is a clan.
Subsets that are not in the form $\bigcup_{j\in J} \C_j$ are not clans either because
each $C_i$ being a strong clan means that no clan overlaps any of them.
Thus, $\M$ is a primitive clan.
\end{proof}

As we can see, the proof of both theorems are similar to the proof of Theorem~\ref{th:closuresmodules} and Theorem~\ref{th:moduletype} but having clans instead of modules. 
The difference between Theorem~\ref{th:moduletype} and Theorem~\ref{th:closuresclansclantype} is the number of strong nodes that are required because when we work with more 
than two equivalence classes is possible to have a primitive node with just three internal nodes:
this is not possible with only two equivalence classes.
For example, we can have coarsest quotient graph that collapses three maximum strong clans, say $\C_0$, $\C_1$ and $\C_2$;
we can reach this situation if they are mutually connected with edges 
in three different equivalence classes because, then, the possible unions of two of them
are not clans, so that the coarsest quotient is primitive.
Such a configuration cannot appear with  only two equivalence classes.

Unless explicitly indicated otherwise,
whenever we talk about closures in the following, we refer to the 
closures described by the set of clan implications. 

It is possible to construct the clan decomposition tree from the closure space lattice getting by the implications obtained from the initial 2-structure. If we take from the closure space just the strong closures, they would be the same sets than the strong clans by Theorem~\ref{th:closuresclansclantype}, 
and the type of the clans are also obtained using Theorem~\ref{th:closuresclansclantype}. However, to give the internal 2-structures of each clan is a more complicated task since we do not have enough information to determine the equivalence class for every edge. 
Thus, so far, we know less about the clan case than about modules.
Advances along this line will hopefully be reported in a future paper.

\subsection{Generalized Gaifman Graphs} 
\label{ssec:ggg}

As we indicated earlier, there are further variations of Gaifman graphs, namely
labeled Gaifman graphs. In the simplest version of labeled Gaifman graphs, the edges are labeled by the multiplicity of the vertices that they connect, that is the number of tuples containing both values: the quantity of co-occurrences of the items into the dataset. 
The simple strategy of having as many equivalence classes as different multiplicities
leads, in practice, to so many equivalence classes that most vertices distinguish most others: no nontrivial clan shows up.
Thus, in a more evolved version, edge labels are obtained from the multiplicities via some sort of discretization process.
We consider two such variations of the labeled Gaifman graphs, based on very
simple discretizations~\cite{DBLP:conf/ida/BalcazarPR18}, and leave for future work a detailed study of the
effect of applying the various existing discretization methods to gather
the diverse multiplicity figures into a sensible quantity of equivalence classes.
Specifically, we report here on the linear and the exponential Gaifman graphs.

In the linear Gaifman graphs, the equivalence classes are determined according to an interval size while, in the exponential Gaifman graphs, the width of the intervals in which the equivalence classes are determined grow in an exponential way. 
More precisely, 
let $c_{x,y}$ be the co-occurrences of the attribute values $x$ and $y$, and let $i$ represent the equivalence class $\E_i$: in a linear Gaifman graph the edge $(x,y) \in \E_i$, if and only if 
$i = \lceil c_{x,y}/n \rceil$, 
being $n$ the assigned interval size; while in an exponential 
Gaifman graph, the edge $(x,y) \in \E_i$, if and only if 
$i = \lceil \log (c_{x,y} + 1) \rceil$, 
where the logarithm is to base two. 
The ceil funtion $\lceil\cdot\rceil$
gives the closest integer larger than or equal to its argument.


We also combine these approaches with the thresholded version of the Gaifman graph. In these variants, all those $c_{x,y}$ below a determined threshold will be consider as if they do not occur often enough and will remain disconnected. It could be possible to work also with upper threshold where those $c_{x,y}$ greater than it are disconnected; we don't study these cases here. 
All these variants require us to work on a 2-structure since, in general, we will have more than two equivalence classes. 


\subsection{Examples of Clan Decomposition: Linear Gaifman Graphs}

\begin{figure}
\centering
\includegraphics[width=2.5in]{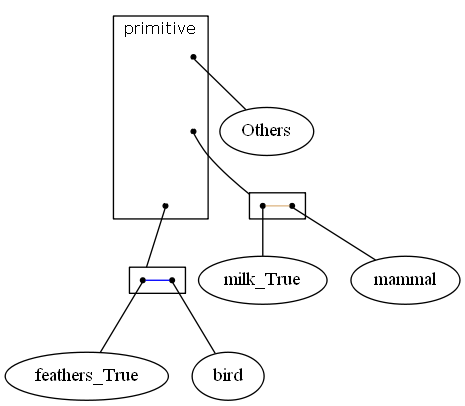}
\caption{Clan decomposition of the Zoo dataset linear 2-structure}
\label{fg:zoolinear}
\end{figure}

In Figure~\ref{fg:zoolinear} we find the result of applying 
the clan decomposition algorithm on the 2-structure given by 
the linear Gaifman graph of the Zoo dataset~\cite{Dua:2019},
with 10 as interval size.
The dataset has seventeen attributes and a total of forty-two attribute values of about one hundred animal species. In the decomposition we find two clans which tell us that mammals feed their offspring with milk and that birds have feathers.


We also work with the linear Gaifman graph decomposition of the Votes dataset, that contains information about the votes for each of the U.S.~House of Representatives Congressmen in 1984 on 16 key votes~\cite{Dua:2019}. The 9 registered types of votes have been simplified into Y, N, and DN (did not vote).

\begin{figure}[h]
\centering
\includegraphics[width=2.5in]{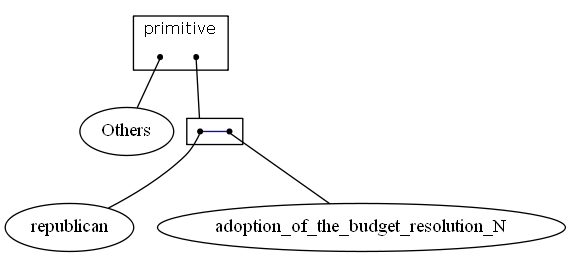}
\caption{Clan decomposition of the Votes dataset linear 2-structure}
\label{fg:voteslinear}
\end{figure}

The Figure~\ref{fg:voteslinear} is the result to apply the decomposition on the linear Gaifman graph with 100 as interval size of those key votes values that appear more than 100 times.  
As you can see we find a clan conform by Republicans and the negative of adoption of the budget resolution, since they are around 141 republicans of 168 votes against adoption of the budget resolution. 
We refrain from deeper analyses here, as our current purpose is just
to illustrate how clan decomposition and Gaifman graphs can be
used for data analysis.

\subsection{Examples of Clan Decomposition: Exponential Gaifman Graphs}

\begin{figure}
\centering
\includegraphics[width=2.5in]{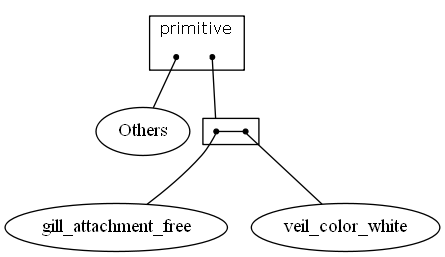}
\caption{Clan decomposition of the Mushroom dataset exponential 2-structure}
\label{fg:mushroomexponential}
\end{figure}

In the Figure~\ref{fg:mushroomexponential} we find the result to apply the clan decomposition algorithm on the 2-structure determined by the exponential Gaifman graph of the Mushroom dataset taking those attribute values that appear more that 2000 times. As we can see most of the attribute values co-occur each other in different ways excepting the items \textit{gill\_attachement\_free} and  \textit{veil\_color\_white } that have the same behavior with the rest of the items. And also we find that they co-occur very often, around 7900 times considering that we have a total of 8000 rows we may say that they appear in the rows almost always, thus most of the mushrooms do not have gills and have a white veil.

\begin{figure}[h]
\begin{center}
\includegraphics[width=1.7in]{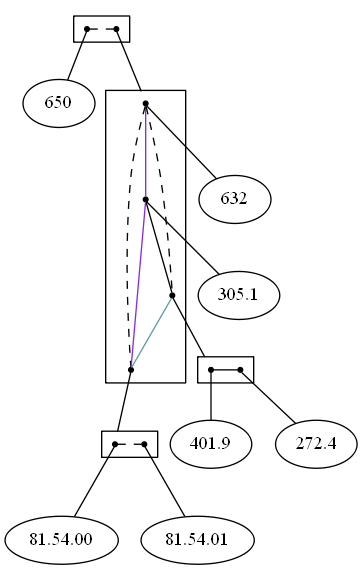}
\end{center}
\caption{Hospitalization graph decomposition.}
\label{fg:hospitalization}
\end{figure}

The next example is part of~\cite{CBMS}, where we apply the clan decomposition method on an exponential Gaifman graph of a hospitalization database. The resulting decomposition is show in Figure~\ref{fg:hospitalization}.  
The hospitalization database has information about hospitalization for the years 2015-2016, each row has information about diagnostics, treatments and conditions of a patient at a fixed date. There is also information related to sex and age of the patient but we do not take this kind of information into account. There are around eighty thousand rows. Whereas the previous datasets were relational, this one is
transactional and serves as an example of ``market-basket-style'' data.

For the 2-structure of our example, we take an exponential Gaifman graph whose vertices are those diagnostics and procedures that appear more than 100 times in the dataset, these seven items: 

\begin{description}
\item [650] Normal delivery, 
\item [632] Missed abortion, 
\item [305.1] Tobacco use disorder, 
\item [401.9] Unspecified essential hypertension, 
\item [272.4] Other and unspecified hyperlipidemia, 
\item [81.54.00] Total knee replacement (left) and 
\item [81.54.01] Total knee replacement (right).
\end{description}

\looseness=1
In the decomposition in the Figure~\ref{fg:hospitalization} we find two small clans: Total knee replacement (left and right) clan and Hypertension and hyperlipidemia clan. 
In the Total knee replacement clan the internal nodes are not related, that is there are no cases where both knees are replaced at once, while in the Hypertension and hyperlipidemia clan its internal nodes are so related, we find around eleven thousand cases of co-ocurrences of them. 
Also, a bigger clan provides us with information as of the approximate 
frequencies with which the small clans show up togheter, for instance 
knee replacements are essentially separate of all the other ailments,
and the root clan tells us that normal delivery does not co-occur with
the other items. The reader may find more detailed investigations on
that dataset in~\cite{CBMS}.

\section{Decomposition algorithm}
\label{sc:decompositionalgorithm}

\looseness=1
Over the years, several algorithms have been studied to 
perform modular decomposition and clan decomposition.
Many of them appear in theoretical publications and
some may fall short of some detail necessary for 
implementation or to ensure whether they work for
clans, beyond modules. Also, the fastest ones are so 
in terms of big-oh notation.

We describe here the algorithm employed in the
implementation of our open-source system to analyze 
data according to 
clan decompositions
\footnote{\url{https://github.com/MelyPic/Gaifman-graphs\_Mod}}.
It is an incremental algorithm, a variation of 
published ones: the vertices of 
the 2-structure are added one by one, and every 
time a vertex is added we get a strong clan 
decomposition tree. We describe afterwards the
similarities and differences with the most similar
among the published algorithms.

Due to incrementality, we can describe the algorithm 
by focusing on the effect of adding one more vertex 
to an existing clan decomposition tree, starting at 
the root; at each recursive call, we have a new item
to add and a 2-structure formed by the coarsest
quotient in the current clan, below which lies the
rest of the decomposition. We call \textit{color of the clan} the equivalence class of the edges in a complete clan.

By comparing the current clan with the new vertex, 
we can clasify the nodes within the clan 
(a coarsest quotient) into three lists:
the list of nodes ``visible with the color of the clan'', 
the list of ``other visible nodes'' and the list of the 
so-called ``nonvisible nodes''. More precisely:

\begin{itemize}
\item The list of nodes visible as the color of the clan can be nonempty only when the root clan is complete, and will contain those nodes whose edges to the new vertex are in the same equivalence class than the internal edges of the clan; this list is empty for primitive clans.   

\item The list of visible nodes contains the nodes into the coarsest quotient of the clan for which an edge can be determined from the new vertex, that is: let $Y_i$ be one of the modules in the coarsest quotient of the current clan and let $x$ be the new vertex, if all the edges from $x$ to the internal nodes of $Y_i$ are in the same equivalence class then we can define an edge from $x$ to $Y_i$. To determine the existence of this edge we use the union-find structure, Subsection~\ref{sec:unionfind}.

\item The list of nonvisible nodes will contain all those nodes in the coarsest quotient of the clan whose internal elements are seen in different ways from the new vertex, which, then, ``does not see'' the clan. Thus we may not assign a single edge color from the new vertex to these nodes, and the clan does not stay in the new decomposition; when we are in this case, the algorithm uses the \textit{split} function from the Algorithm~\ref{alg:splitalgorithm}, it is explained in detail in the Subsection~\ref{sec:split}. 
\end{itemize}

\subsection{Union-Find Structure}
\label{sec:unionfind}
We use the union-find data 
structure
\footnote{~\url{https://en.wikipedia.org/wiki/Disjoint-set\_data\_structure}}
to get the defined edges between vertices, between clans and between vertices and clans. In fact, to implement these functions we rely directly on the pseudocode that appears in~\cite{DBLP:books/daglib/0023376}.
  
The edges between vertices are initialized according to the original 2-structure.
Let $x$ and $y$ be two vertices in the graph, the function \textit{Find} returns the equivalence class for the edge that connects them. Thus edges in the same equivalence class will be return the same value for the function \textit{Find}. 

In this way, when a new clan is obtained, the edges to it from the other clans and vertices (if they exist) are generated using the \textit{Makeset} function and using the \textit{Union} function those edges are added to their corresponding equivalence class determined by the \textit{Find} function.
So, to know if an edge is defined it is enough to call its \textit{Find} function.

%
%
     %
\subsubsection{Complexity:}
We implement the Union-Find structure using the union-by-rank heuristic with a running time of $O(m \log n)$, where $m$ is the number of \textit{MakeSet}, \textit{Union} and \textit{Find} operations and $n$ is the number of \textit{MakeSet} operations~\cite{DBLP:books/daglib/0023376}.



\subsection{Pack Function}
\label{sec:pack}

The pack function is executed in order to determine the edges to a clan. It is applied once a clan is obtained or when a new element is added to an existing clan.
The purpose of this is to help us to obtain in a linear process the lists of nodes ``visible with the color of the clan'', the list of ``other visible nodes'' and the list of the ``nonvisible nodes'' as defined above. 
We use Algorithm~\ref{alg:packalgorithm} for this function.  

\subsubsection{Complexity:}
Let $n$ be the total number of vertices of the 2-structure $\G$, let $k$ be the total number of vertices in the clan $C$ and let $|CQ_C|$ be the total number of elements in the coarsest quotient of $C$, the algorithm takes $(n - k) \times |CQ_C|$ time. That is because the step \textbf{for} in line 5 runs for $|V_{\G-C}|$ loops; as $V_C$ is the list of vertices in $\G$ not in $C$, its length is $n-k$. The \textbf{while} in line 10 takes time at most $|CQ_C|$.

The worst case for the coarsest quotient is to have just single items, that is $|CQ_C|= k$, in this way, the algorithm takes $(n-k) \times k = nk - k^2$ times, where $k \leq n$.


\begin{algorithm}
\caption{Algorithm to pack a clan of a 2-structure}
\begin{algorithmic}[1]
\STATE \textbf{Input:} A 2-structure $\G$ and a clan $C$ on it.
\STATE\textbf{Output:} The edges from/to the clan $C$ using the Union-Find data structure.
\STATE Let $CQ_C$ be the list of elements into the coarsest quotient of $C$.
\STATE Let $V_{\G-C}$ be the list of vertices in $\G$ not in $C$.
\FORALL{$v$ in $V_{\G-C}$}
	\STATE InitialEdge = $v,CQ_C[0]$
	\STATE InitialClass = Find(InitialEdge)
	\STATE SameClass = True
	\STATE j = 1
	\WHILE{SameClass and $j < len(CQ_C)$}
		\IF{InitialClass == Find($v,CQ_C[j]$)}
			\STATE j+=1
		\ELSE
			\STATE SameClass = False			
		\ENDIF		
	\ENDWHILE
	\IF{SameClass}
		\STATE NewEdgeTo = $v,C$
		\STATE NewEdgeFrom = $C,v$
		\STATE MakeSet(NewEdgeTo)
		\STATE MakeSet(NewEdgeFrom)
		\STATE Union(InitialEdge,NewEdgeTo)
		\STATE Union(InitialEdge,NewEdgeFrom)		
	\ENDIF
\ENDFOR
\end{algorithmic}
\label{alg:packalgorithm}
\end{algorithm}

\subsection{Split Function}
\label{sec:split}
Assume the vertex $x$ sees differently some of the internal nodes of some clan $Y$, thus we have to split $Y$. 
To split a clan means traversing clans below it in the tree sufficiently
deep so as to finding well-defined edges to $x$. 
Let $Y_0, Y_1, \ldots , Y_n$ be the elements of the coarsest quotient of $Y$, the result of split $Y$ with respect to $x$ includes those $Y_i$ that have an edge defined from $x$; then,
any $Y_j$ on $Y$ ``not seen'' by $x$ must be recursively split. The algorithm to split a node is the Algorithm~\ref{alg:splitalgorithm}, based on the Theorem~\ref{th:split}.

\subsubsection{Complexity:}
The split function makes a recursive call when any of the elements in the corresponding coarsest quotient is not distinguishable, does not matter the clan type (lines 9 and 18 of Algorithm~\ref{alg:splitalgorithm}). thus the worst case is to do the split of all clans in the decomposition tree, unless all those elements are leaves. 
The clan decomposition tree with most clans is that with just complete clans, whose coarsest quotients have two elements: assume, one of the elements in all the coarsest quotient is a leaf, in this way we get the decomposition tree as deeply as possible, having a total of $n-1$ clans for a 2-structure with $n$ vertices, as the last clan have only leaves, the maximum clans where we can apply the split function is $n-2$; 
while, if both of the elements in all the coarsest quotient are clans, for a 2-structure with $n$ vertices we will apply the split function in a total of $\sum_{i=1}^{k-1} 2^{i} + (n-2k)$ times, where $k = \left\lfloor \log_{2} n\right\rfloor$. 

Assume  $n-2 > \sum_{i=1}^{k-1} 2^{i} + (n-2k)$ where $k = \left\lfloor \log_{2} n\right\rfloor$:
$$n > 2 + \sum_{i=1}^{k-1} 2^{i} + (n-2k)$$
$$0 > 2 + \sum_{i=1}^{k-1} 2^{i} - 2k$$
$$0 > 1 + \sum_{i=0}^{k-1} 2^{i} - 2k$$
$$0 > 1 +  2^{k} - 1 - 2k$$
$$0 > 2^{k} - 2k,$$ a contradiction.

 Thus, for a 2-structure with $n$ vertices the split function take at most $\sum_{i=1}^{k-1} 2^{i} + (n-2k)$ where $k = \left\lfloor \log_{2} n\right\rfloor$.
As  $k = \left\lfloor \log_{2} n\right\rfloor$:
$$\sum_{i=1}^{k-1} 2^{i} + (n-2k) < 1+ \sum_{i=1}^{k-1} 2^{i} +(n-2k) $$

$$\sum_{i=1}^{k-1} 2^{i} + (n-2k) < \sum_{i=0}^{k-1} 2^{i} +(n-2k) $$

$$\sum_{i=1}^{k-1} 2^{i} + (n-2k) < 2^{k} - 1 + n-2k $$

$$\sum_{i=1}^{k-1} 2^{i} + (n-2k) < n - 1 + n-2k $$

$$\sum_{i=1}^{k-1} 2^{i} + (n-2k) < 2n -2k -1 $$

Thus, the split function runs in time $O(n)$.

\begin{algorithm}
\caption{Algorithm to split a clan from a node}
\begin{algorithmic}[1]
\STATE \textbf{Input:} A clan $C$ and a node $n$.
\STATE\textbf{Output:} $C_v$, the set of maximal strong clans below $C$ visible from $n$.
\STATE Let $C_v$ be an empty set.
\IF{$C$ is primitive}
	\FORALL{$\M$ in the coarsest quotient of $C$}
		\IF{$\M$ is visible from $n$}
			\STATE Add $\M$ to the maximal strong clans in $C_v$.		
		\ELSE
			\STATE Split $\M$ from $n$ (Algorithm~\ref{alg:splitalgorithm} with $\M$ and $n$ as parameters), add each maximal strong clan in the result to the set of maximal strong clans in $C_v$.
		\ENDIF
	\ENDFOR	
\ELSE
	\STATE Let $C_i$ be an empty set for each different equivalence class $i$ (color).
	\FORALL{$\M$ in the coarsest quotient of $C$}
		\IF{$\M$ is visible from $n$}
			\STATE Add $\M$ to $C_i$, being $i$ the way in which $n$ sees $\M$.			 		
		\ELSE
			\STATE Split $\M$ from $n$ (Algorithm~\ref{alg:splitalgorithm} with $\M$ and $n$ as parameters), add each maximal strong clan in the result to the set of maximal strong clans in $C_v$.
		\ENDIF
	\ENDFOR
	\FORALL{$C_i$}
		\IF {$C_i$ has more than one maximal strong clan}
			\STATE Add $C_i$ (as a single clan) to the set of maximal strong clans in $C_v$.
		\ELSE
			\STATE Add the unique $\M$ in $C_i$ to the set of maximal strong clans in $C_v$.
		\ENDIF				
	\ENDFOR
\ENDIF
\STATE Return $C_v$.		
\end{algorithmic}
\label{alg:splitalgorithm}
\end{algorithm}

\subsection{Clan Decomposition Algorithm}

As an initial case to the algorithm, we have two cases. One of them is to add the new vertex to an empty tree, in this case the vertex becomes a complete clan with just the vertex in its coarsest quotient. The other case is to add the new vertex to a tree encompassing just one item; then, the root is a singleton clan with vertex $x$ and, in this case, the new vertex is added to form a complete clan with two singletons: the new vertex and $x$. The color (equivalence class) of the edge is unambiguous.

Assume we have a non-empty strong clan tree and a new node to add. We have to look for its position into the tree, starting from the root, going over the strong tree as deep as necessary until finding its place: a clan, represented as its coarsest quotient graph so each node there corresponds to a subtree decomposing the corresponding maximal strong clan; then we may walk towards the leaves recursively. We will refer to the clan in which we are as current clan. 

If we are not in any of the initial cases we follow the next steps:

\begin{enumerate}
\item If the current clan is a complete clan, the new vertex can:
\begin{enumerate} 
\item Become one member of the clan, preserving the type of the clan: if the new node sees all the internal nodes of the current clan with the same color as the color of the clan, then the new node is added as a maximal strong trivial clan into the coarsest quotient of the current clan and the type of clan continues being complete. No recursive call is made. 
\item Generate a subclan: The new node sees at least one but not all the internal nodes with the same color as the color of the clan, those elements that are not seen by the color of the clan are removed from the coarsest quotient of the current clan and moved to the coarsest quotient of a new complete clan, sibling to the nodes seen by the color of the clan. Thus, this new complete subclan and the nodes seen by the color of the clan will conform the coarsest quotient of the current clan. The new vertex is recursively added to the new clan. 
\item Generate a superior clan: The internal nodes are in the list of visible nodes and are seen by $x$ all as the same color but different from the color of the clan. Then, the current clan and the new vertex will be maximal strong clans into the coarsest quotient of a superior complete clan of size 2. No recursive call is made. 
\item Become one member of the clan, changing the type of the clan to primitive: When not all the nodes in the clan are seen by the same color and none of them are seen by the color of the clan. Thus, we add the new vertex to the coarsest quotient of the current clan and:
	\begin{enumerate} 
	\item The nodes in the list of visible nodes are grouped into clans according how they are seen from the new vertex and,
	\item The nodes in the nonvisible list are split. 
	\end{enumerate}
\end{enumerate}


\item If the current clan is a primitive clan, the new vertex can:
\begin{enumerate} 
\item Become added (recursively) to the subclan of one of the nodes inside of the coarsest quotient of the current clan. This happens when the new vertex and one of the internal nodes see the remaining nodes in the same way. Theorem~\ref{th:clanwithone} supports this step by proving that the branch for the recursive call is always
uniquely defined.
\item Generate a superior clan: if all of the internal nodes are in the list of visible nodes and are seen by the same color. Then, the current clan and the new vertex will configure the coarsest quotient of a superior complete clan of size 2. 
\item Become one member of the coarsest quotient of the current clan, preserving the type of the clan: when not all the nodes in the list of visible nodes are seen by the same color. Thus, the new vertex is added to the coarsest quotient of the current clan and the nodes in the nonvisible list are split. 
\end{enumerate} 
\end{enumerate}

The general algorithm to process a graph is Algorithm~\ref{alg:decompalgorithm}. It uses the Algorithm~\ref{alg:accordtype} to add a new vertex.

In turn the Algorithm~\ref{alg:accordtype} uses the Algorithm~\ref{alg:splitalgorithm} to split a node 
and the Algorithm~\ref{alg:packalgorithm} to determine the existing edges from the new clan to the remaining vertices.

\begin{algorithm} 
\caption{Algorithm to obtain a strong clan decomposition tree from a 2-structure}
\begin{algorithmic}[1]
\STATE \textbf{Input:} A 2-structure. 
\STATE \textbf{Output:} Strong clan decomposition tree.
\STATE Let the current strong clan decomposition tree be an empty clan.
\STATE Let $\U$ be the set of vertices in the 2-structure.
\FORALL {$v \in \U$}
	\STATE Current strong clan = root clan of the current strong clan decomposition tree.
	\STATE Current strong clan decomposition tree = Apply the Algorithm~\ref{alg:accordtype} with $v$, the current strong clan and the respective 2-structure induced by them as parameters.
\ENDFOR
\end{algorithmic}
\label{alg:decompalgorithm}
\end{algorithm}

\subsubsection{Complexity of Algorithm~\ref{alg:decompalgorithm}:} Let $n$ be the total number of vertices of a 2-structure, the run time of Algorithm~\ref{alg:decompalgorithm} is $\sum_{i=1}^{n} i$ since Algorithm~\ref{alg:accordtype} is linear. Thus, the Algorithm~\ref{alg:decompalgorithm} runs in $O(n^2)$

\begin{figure}
\begin{algorithm}[H]
{\small
\caption{Algorithm to get the strong clan decomposition tree when a vertex is added to a specific clan.}
\begin{algorithmic}[1]
\STATE \textbf{Input:} A vertex $v$ to be added, a clan $C$ and the 2-structure induced by the vertices on $C$ plus $v$.  
\STATE \textbf{Output:} Strong clan decomposition tree. 
\STATE{Let $L_n$ be the list of nonvisible maximal strong clans in the coarsest quotient of $C$ from $v$},
\STATE{Let $L_v$ be the list of visible maximal strong clans in the coarsest quotient of $C$ from $v$},
\STATE{Let $L_c$ be the list of visible, as the color of the clan, maximal strong clans in the coarsest quotient of $C$ from $v$.}
\IF{$C$ is empty or its coarsest quotient has just one element}
		\STATE Add $v$ to the coarsest quotient of $C$.	
		\STATE Assign the type of $C$ as complete.	
\ELSIF{the type of $C$ is complete}
	\IF {$L_c$ is equal to the maximal strong clans in the coarsest quotient of $C$}
			\STATE Add $v$ to the coarsest quotient of $C$.			
	\ELSIF{$|L_c| \geq 1$}
		\STATE Let $C_{aux}$ be an auxiliary complete clan.
		\FORALL {maximal strong clans $\M$ not in $L_c$}
			\STATE Remove $\M$ from the coarsest quotient of $C$.
			\STATE Add $\M$ to the coarsest quotient of $C_{aux}$.
		\ENDFOR	
		\STATE Add $C_{aux}$ to the coarsest quotient of $C$.
		\STATE Add $v$ to the clan $C_{aux}$. (Algorithm~\ref{alg:accordtype} with $v$, $C_{aux}$ and the respective 2-structure as parameters).
		\STATE Pack the coarsest quotient of $C_{aux}$. (Algorithm~\ref{alg:packalgorithm}) 
		\ELSIF {$L_v$ is equal to the maximal strong clans in the coarsest quotient of $C$ and all of them are seen in the same way from $v$}
			\STATE Copy the elements to the coarsest quotient of $C$ to the auxiliary complete clan $C_{aux}$, and remove them from $C$. 
			\STATE Add $C_{aux}$ and $v$ to the coarsest quotient of $C$.
		\ELSE
			\STATE Change the type of the clan $C$ to primitive.
			\FORALL{maximal strong clans $\M$ in $L_n$}
				\STATE Remove the $\M$ from the coarsest quotient of $C$.				
				\STATE Split $\M$ from $v$ (Algorithm~\ref{alg:splitalgorithm} with $\M$ and $v$ as parameters), add each maximal strong clan in the result to the coarsest quotient of $C$.
			\ENDFOR				
	\ENDIF
	\ELSE
		\IF {there is one maximal strong clan $\M$ in the coarsest quotient of $C$ that sees the remaining maximal strong clans in the same way than $v$ does}
			\STATE Add $v$ to the clan $\M$ (Algorithm~\ref{alg:accordtype} with $v$, $\M$ and the respective 2-structure as parameters).
		\ELSIF{$L_v$ is equal to the maximal strong clans in the coarsest quotient of $C$ and all of them are seen in the same way from $v$}
				\STATE Copy the elements to the coarsest quotient of $C$ to the auxiliary primitive clan $C_{aux}$, and remove them from $C$. 
				\STATE Change the type of the clan $C$ to complete.
				\STATE Add $C_{aux}$ and $v$ to the coarsest quotient of $C$.
		\ELSE
			\STATE Add $v$ to the coarsest quotient of $C$
			\FORALL{maximal strong clans $\M$ in $L_n$}
				\STATE Remove the $\M$ from the coarsest quotient of $C$.				
				\STATE Split $\M$ from $v$ (Algorithm~\ref{alg:splitalgorithm} with $\M$ and $v$ as parameters), add each maximal strong clan in  the result to the coarsest quotient of $C$.
			\ENDFOR		
		\ENDIF
\ENDIF
\STATE Pack $C$ (Algorithm~\ref{alg:packalgorithm} with $\G$ and $C$ as parameters).
\STATE Return $C$
\end{algorithmic}
\label{alg:accordtype}
}
\end{algorithm}
\end{figure}

\subsubsection{Complexity of Algorithm~\ref{alg:accordtype}:} 
Let $|CQ _C |$ be the total number of elements in the coarsest quotient of $C$ and let $|L_v|$ be the length of the list $L_v$, the \textbf{for} on line 20 takes $|CQ_C|-|L_v|$ times.
The \textbf{for}'s involve for the copy of the coarsest quotient elements on lines 28 and 41 takes $|CQ_C|$ times each one. 
While the \textbf{for}s in lines 32 and 45 takes $|L_n|$ plus the time of the Split function that is linear on the number of vertices involve.Thus, in the worst case $|L_n|=|CQ_C|$ so that an element within the coarsest quotient of $C$ is not distinguished, it must have at least two vertices being the maximum size of $|CQ_C|= \frac{n}{2}$. As each element into the coarsest quotient has two vertices, the split function will have 2 as running time, having all the process $n$ as running time.  

In this way the complexity of Algorithm~\ref{alg:accordtype} is linear on the number of vertices.

\subsection{Algorithms~\ref{alg:splitalgorithm},~\ref{alg:decompalgorithm} and~\ref{alg:accordtype}: Correctness and Examples}
\label{ssec:CorrectnessAlgorithms}

Let $\C \subseteq \U$ be a clan to which we attempt at adding $x \in \U$. We assume $|\C| \geq 2$ as the base cases are directly correct by definition. 
Let $\M$ be the coarsest quotient of $\C$, which can be either primitive or complete; it consists of maximal strong clans $\M = \{ \C_i | i \in I\}$ for some index set $I$, with $\C_i\subseteq \U$, each $\C_i$ is handled
through its own coarsest quotient $\M_i$ like $\C$ is
handled through $\M$. We allow for the
slight abuse of notation $(\C_i,\C_j)\in\E_p$ to refer to
all edges with one endpoint within $\C_i$ and another 
within $\C_j$ being in $\E_p$. Since they are clans, 
such $p$ is well-defined.

If $\M$ is a complete clan, then $p_{\M}$ is the ``color of the clan'', 
that is, the index of the equivalence class of its edges, $\E_{p_{\M}}$; we often omit the subindex $\M$ when it is clear from the context. 
We denote $J\subseteq I$ the (indices of) maximal strong clans in it that are seen from $x$ with color $p_{\M}$: $J = \{ i\in I | \forall z \in\C_i (x,z)\in\E_p \}$.
This is the formalization of the ``list of nodes
visible with the color of the clan''.

Likewise, $\C_i$ being visible from $x$ is formalized as
$\exists q \forall z\in\C_i ((x,z)\in\E_q)$ and its negation,
which says that adding $x$ requires to split $\C_i$, is 
$\forall q \exists z\in\M_i ((x,z)\notin\E_q)$.

Since maximal strong clans in $\M$ are proper subclans of $\C$,  $|I| \geq 2$.
Additionally, it is known that if $\M$ is primitive then
every proper subclan $\C'\subset \C$ is included
in one of the maximal strong subclans: $\exists i (\C'\subseteq\C_i)$.

\begin{definition}
Given a clan $\C\subset\U$,
a proper subclan of $\C'\subset\C$, 
and a vertex $x\notin\C$,
we say that $x$ is like $\C'$
in $\C$ if $x$ sees the rest of $\C$ in the same way as $\C'$:
for all $y \in \C'$, and for all $z\in(\C\setminus\C')$,
the edges $(x,z)$ and $(y,z)$ are equivalent.
\end{definition}

The following theorems are focused on identifying the components of $\M'$, the desired coarsest quotient of $\C\cup\{x\}$, for the cases (1)(b), (1)(d), (2)(c) and (2)(a) of the Algorithm~\ref{alg:accordtype}.
The correctness of the rest of the cases is easier to argument.

For the case (1)(b) above, we have the following theorem.

\begin{theorem}
\label{prop:InotJnonempty}
If $\M$ is complete and $\emptyset \neq J \neq I$, then 
$\M' = \{ \C_j | j\in J \} \cup \{ \{ x \} \cup\bigcup_{j\notin J} \C_j \}$. 

\end{theorem}

\begin{proof}
Items inside each $\C_j$ with $j\in J$, which is a clan of $\C$, 
are not distinguishable from either other $\C_j$'s or $x$;
thus, they are clans of $\C\cup\{x\}$. By the same token,
so is $\{ \{ x \} \cup\bigcup_{j\notin J} \C_j \}$. 
Assume they are not strong clans. Overlaps of other 
clans within $\C$ with the $\C_i$'s cannot exist
because $\M$ is the coarsest quotient of $\C$. Hence,
a hypothetical overlapping clan must include $x$ and some $y\in\C_j$
for some $j\in J$, but not all $z\in \bigcup_{j\notin J} \C_j$.
But, then, $(y,z)\in\E_p$ and $(x,z)\notin\E_p$ so it
is not a clan.
Thus, they are strong clans. Larger clans 
cannot exist either as, then, ignoring $x$
would lead to larger clans already in $\M$.
\end{proof}

In the next theorem, the complete case corresponds to (1)(d) above 
and the primitive case is
(2)(c). 
The sequence of maximal strong clans corresponds to
the successive recursive calls of Algorithm~\ref{alg:splitalgorithm}.

\begin{theorem}
\label{th:split}
If $\M$ is primitive and $x$ is not like any $\C_i$, 
or if $\M$ is complete with $J = \emptyset$, 
and besides, in either case, $\forall q \exists z\in\C (x,z)\notin\E_q$,
then $\C'\in\M'$ if and only if $\C' = \{ x \}$ or
there is a sequence 
$\C = \C'_0 \supset \C'_1 \supset \cdots \supset \C'_n \supset \C'$,
where: 
\begin{enumerate}
\item
each $\C'_{i+1}$ is a maximal strong clan in the coarsest
quotient of $\C'_i$ for $i < n$ (holds vacuously if $n = 0$),
\item
$x$ does not split $\C'$,
\item
$x$ splits each $\C'_i$ for $i \leq n$, and either
\begin{enumerate}
\item
the coarsest quotient of $\C'_n$ is primitive and
$\C'$ is one of its components or
\item
the coarsest quotient of $\C'_n$ is complete and
there is $q$ such that 
$\C'$ is the union of all those components
of the coarsest quotient of $\C'_n$ that
are connected to $x$ with edges of class $\E_q$.
\end{enumerate}
\end{enumerate}
\end{theorem}

Note that, 
for some such $\C'$, $n$
can turn out to be $0$.

\begin{proof}
We study first the case of $\{ x \}$.
Let $\C'$ be the maximal strong clan $\C'\in\M'$
that has $x\in\C'$. Assume $|\C'|\geq 2$.

In case $\M$ is complete, say with color $p$, and
with $J = \emptyset$, 
use $|\C'|\geq 2$ to fix $z\in\C'$ such that $z\neq x$,
and let $\C_{i_z}\in\M$ such that $z\in\C_{i_z}$. 
Since $|\M|\geq 2$, there is at least some
other $\C_{i_y}\in\M$, $i_y \neq i_z$; and
some $y\in\C_{i_y}$ must have $(x,y)\notin\E_p$
as, otherwise, $i_y\in J$.
Then, $y$ sees differently $x$ and $z$
because $(y,z)\in\E_{p}$, the color of $\M$, which
contradicts the fact that $\C'$ is a clan.

Now, in case $\M$ is primitive, since $|\C'|\geq 2$, 
the set $\C' \setminus \{ x \}$ 
is nonempty and, hence, also a clan in $\C$:
there is $i$ such that $\C' \setminus \{ x \}\subseteq\C_i$,
and the maximality of $\C'$ implies equality:
$\C' \setminus \{ x \} = \C_i$. Then, the other
maximal strong clans cannot distinguish $\C_i$ from
$x$ as, otherwise, $\C'$ is not a clan, and therefore
$x$ is like $\C_i$, in contradiction with the hypothesis.

\looseness=-1
Thus, in both cases, $|\C'| = 1$ and necessarily $\C' = \{ x \}$.
That is, the single clan in $\M'$ containing $x$
is the singleton $\{ x \}$.

\looseness=-1
We move on to the case where $x\notin\C'$,
which implies $\C'\subseteq \C$. It has to
be a proper subset, $\C'\subset \C$, because
$\C$ in full is not a clan in $\C\cup\{x\}$:
otherwise we would contradict the hypothesis
that $\forall q \exists z\in\C (x,z)\notin\E_q$
(equivalently, $x$ splits $\C$).

Assume first $\C'\in\M'$. Together with 
$\{ x \}\in\M'$ that we have already argued, 
this implies that there is $q$ such that 
$\forall z\in\C' ((x,z)\in\E_q)$. Fix that $q$.
Also, $\C'$ must be a clan in $\C$ as, otherwise,
it cannot be a clan in the larger set $\C\cup\{x\}$
and, by the same reason, $\C'$ is not split by $x$.

\looseness=-1
We construct the claimed sequence inductively.
The basis is, of course, $\C'_0 = \C$.
Note that we just argued that $x$ splits it.

Suppose the construction has proceeded up to
some $\C'_i \supset \C'$ where $x$ splits $\C'_i$,
the inclusion being proper because $x$ does 
not split $\C'$.
The coarsest quotient of $\C'_i$
can be either primitive or complete.

If the coarsest quotient of $\C'_i$
is primitive, then there is $\C'_{i+1}$ in the coarsest
quotient of $\C'_i$ such that 
$\C'_{i+1} \supseteq \C'$.
We check whether all the edges from
$x$ to $\C'_{i+1}$ are equivalent. If so, then
$\C'_{i+1}$ is a strong clan in $\C\cup\{x\}$ and, 
by maximality of $\C'$, it must be $\C'_{i+1} = \C'$,
so that $\C'$ is one of the components of the
coarsest quotient of $\C_i$ and the 
construction stops with $n = i$.
Otherwise, $x$ splits $\C'_{i+1}$ and the construction
has proceeded one further step.

Alternatively, if the coarsest quotient 
of $\C'_i$ is complete, we consider its
decomposition $\{ \C''_j | j \in I' \}$
for some index set $I'$.
By the properties of complete clans, $\C'$ being
a clan in $\C$ is either a proper subset of
some $\C''_j$, or a union of one or more of them
(but not all since $\C'\subset \C'_i$ properly).
In the latter case, we have the claimed statement
by finishing the construction with $n = i$
and letting $q$ be the color of the edges from $\C'$
to $x$; maximality of $\C'$ implies that all the components
having that edge color in the connection to $x$ are
included in $\C'$.

Finally, in the former case, $\C'\subset \C''_j$ for $j\in I'$,
note that if $\C''_j$ were not split by $x$, 
$\C''_j$ being a clan of $\C$, 
then it would be a clan 
in $\C \cup \{ x \}$, contradicting the
maximality of $\C'$. Thus, $\C''_j$ 
must be split by $x$ and the construction
goes on, taking this $\C''_j$ as $\C'_{i+1}$.

The construction cannot run forever because $\C$ is finite.


Conversely, we must prove that every set $\C'$ for which 
a chain as given exists is in $\M'$: we have that
$\C = \C'_0 \supset \C'_1 \supset \cdots \supset \C'_n \supset \C'$
(each $\C'_i$ a maximal strong clan in the previous one,
each split by $x$, but not $\C'$)
and that $\C'$ is either one of the components of a 
primitive coarsest quotient of $\C_n$,
or a union of one or more components of a complete
coarsest quotient of $\C_n$, namely, those that see
$x$ with color $q$.

We need to see that $\C'$ is a clan (in $\C \cup \{ x \}$), 
strong, and maximal. When the coarsest quotient of $\C_n$ 
is primitive with $\C'$ one of its components, the chain
of maximal strong clans keeps $\C'$ a strong clan all the
way up; besides, these are the only strong clans that can
contain $\C'$. Additionally, $\C$ is also maximal 
because $x$ splits all the other clans along the way,
so no superset of $\C$ is a strong clan of $\C \cup \{ x \}$.

When the coarsest quotient of $\C_n$ is complete, 
$\C'$ is a union of components 
in that coarsest quotient;
therefore, it is a clan in $\C$. 
Also, $x$ does not split $\C'$ either
because all those components are connected to $x$
by edges of color $q$, so $\C'$ is a clan in $\C \cup \{ x \}$.

To see that it is a strong clan, observe that a hypothetical clan
including part of $\C'$ plus some proper part of 
another component of $\C_n$ would overlap that
component, which cannot be the case because all
these components are strong; and cannot include any
other component as a whole, because all those connected
by color $q$ were already taken in for $\C'$ so any other one
is distinguished from $\C'$ by $x$. Additionally, if 
a hypothetical clan overlaps $\C$ by including vertices
from outside $\C_n$, then it would overlap $\C_n$, which
cannot be the case either as $\C_n$ is a strong clan.

The argument for maximality is similar: 
now, the hypothetical clan includes all of $\C'$
instead of a part of it, plus additional vertices,
and we argue like before:
if these vertices come from $\C_n$, either
they are a proper subset of another component, 
and then that component would not be strong, or
they include totally other components, and then 
$x$ distinguishes them and is not a clan;
and if they come from outside $\C_n$, then they 
would overlap $\C_n$, which cannot be the case;
that is because not all the components of
$\C_n$ are taken into $\C'$ since we know 
$\C' \subset \C'_n$ properly and the components
left outside $\C'$ are distinguished from $\C'$ by $x$.
\end{proof}

Regarding case (2)(a) above, the fact that needs
specific analysis is the reason why one subclan
to recurse on is unambiguously identified and will
lead to a correct outcome.

\begin{theorem}
\label{th:clanwithone}
Let $\M$ be a primitive coarsest quotient of a clan $\C$ 
and let $x\notin\C$. Then there exists at most one 
maximal strong subclan $\C_i\in\M$ such that $x$ 
is like $\C_i$ in $\C$. If there exists one such
$\C_i$, then $\C_i \cup \{ x \}$ is a maximal strong
subclan of $\C \cup \{ x \}$.
\end{theorem}

\begin{proof}
Consider $\C_i\in\M$ and $\C_j\in\M$ with $i\neq j$,
and suppose $x$ is like $\C_i$. Since $\M$ is primitive,
there is a third clan $\C_k\in\M$ that distinguishes $\C_i$
from $\C_j$. As $x$ is like $\C_i$, the edges connecting
$x$ to $\C_k$ are equivalent to those connecting $\C_i$
to $\C_k$, hence they are not equivalent to those connecting
$\C_j$ to $\C_k$; this implies that $x$ is not like $\C_j$.

If such $\C_i$ is available, let $\C' = \C_i \cup\{ x \}$:
by the choice of $i$, it is a clan; it must be a strong
clan because, as just argued, $x$ cannot form a clan with
any other $\C_j$ and no clan of $\C$ can overlap $\C_i$,
which was already strong. Likewise, any proper superset
of $\C'$ becomes a proper superset of $\C_i$ once $x$
is removed, and cannot be a clan due to the maximality of $\C_i$.
\end{proof}


%

\subsubsection{Examples}
Let us see a couple of examples to illustrate the algorithms.

\begin{example}
As an example of how the Algorithm~\ref{alg:accordtype} works, we apply them on the graph of Figure~\ref{fg:ExampDecomp}. It is displayed there as a 2-structure with two equivalence classes.
We keep adding vertices in the standard alphabetical order.

\begin{figure}[ht]
\centering
\includegraphics[width=2.95in]{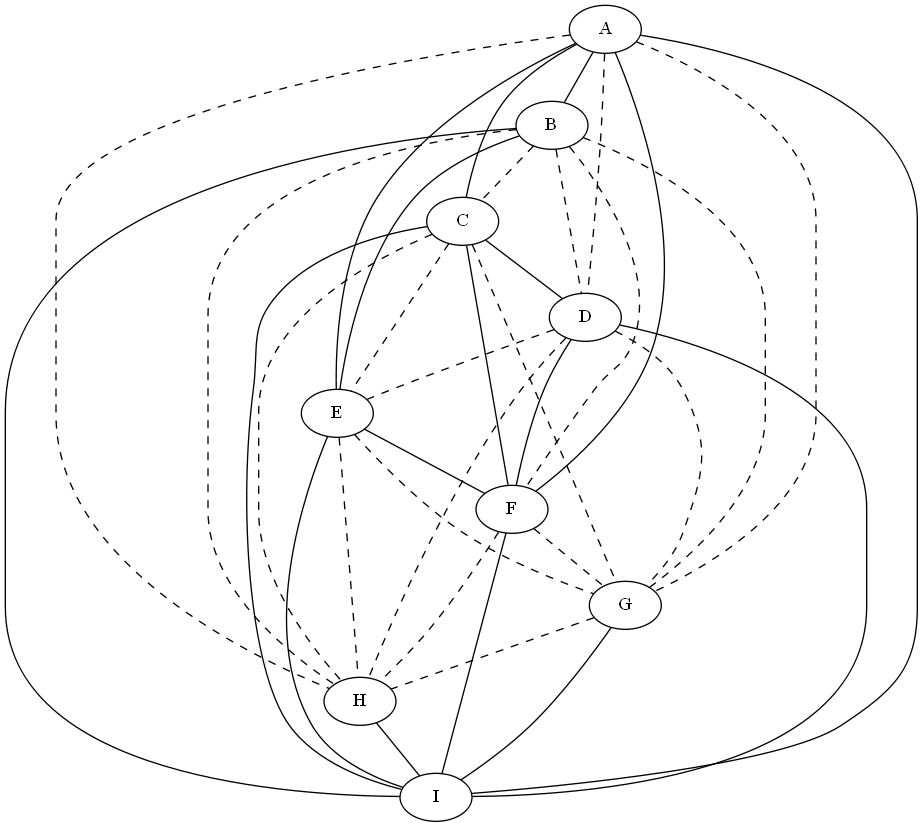}
\caption{Example: Graph to apply the decomposition Algorithm~\ref{alg:accordtype}.}
\label{fg:ExampDecomp}
\end{figure}

At the beginnig we add vertex $a$ to an empty clan decomposition tree, getting a tree with just one clan. In the next step we add the vertex $b$ to it, having as a result a clan with two maximal strong clans in its coarsest quotient (a complete clan). Both steps are initial cases, they are shown in Figure~\ref{fg:ExampDecomp_0}. 

\begin{figure}[ht]
\centering
\includegraphics[width=1in]{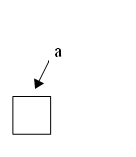}
\includegraphics[width=.57in]{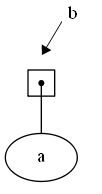}
\includegraphics[width=1in]{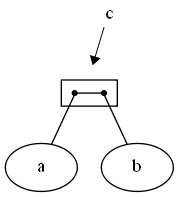}
\caption{Initial cases for Algorithm~\ref{alg:accordtype}}
\label{fg:ExampDecomp_0}
\end{figure}

When we add $c$ to this clan, we are in the case $1.(b)$ since $a$ is seen from $c$ by the color of the clan but $b$ is not. Thus, the maximal strong clans that are not seen from the color of the clan are moved to another sibling clan, in this case containing $b$, and $c$ is added to it, in this case we add $c$ to a complete clan with just $b$ in its coarsest quotient (the initial case). All the process is shown in Figure~\ref{fg:ExampDecomp_1_1b}.  

\begin{figure}[ht]
\centering
\includegraphics[width=1in]{02_Insert_c.png}
\includegraphics[width=1in]{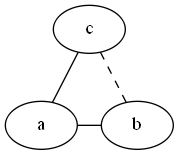}
\includegraphics[width=1in]{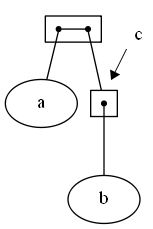}
\includegraphics[width=1in]{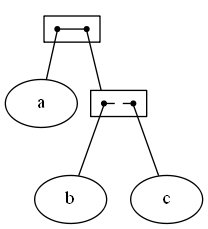}
\caption{Applying 1.(b) from Algorithm~\ref{alg:accordtype}}
\label{fg:ExampDecomp_1_1b}
\end{figure}

Now, we have in the root a complete clan and the node $d$ is added to it. We find the new node may not see one of the maximal strong clans in the coarsest quotient of the root, the case $1.(d)$, center of Figure~\ref{fg:ExampDecomp_2_1d}. Thus, the type of the clan is changed to primitive and the maximal strong clans that are seen from $d$ in the same way will conform a new coarsest quotient while those nodes that are not seen from $d$ are split; in this case we do not have any new coarsest quotient and the maximal strong clan conformed by $b$ and $c$ is split, getting as result the decomposition tree shown in the right of Figure~\ref{fg:ExampDecomp_2_1d}. 



\begin{figure}[ht]
\centering
\includegraphics[width=1.2in]{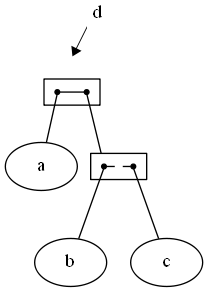}
\includegraphics[width=1in]{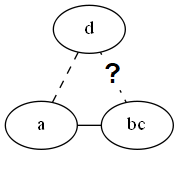}
\includegraphics[width=.8in]{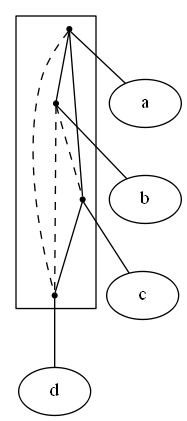}
\caption{Applying 1.(d) from Algorithm~\ref{alg:accordtype}}
\label{fg:ExampDecomp_2_1d}
\end{figure}

In the next step we add the vertex $e$ to a primitive clan root, left of Figure~\ref{fg:ExampDecomp_3_2a}. We find that there is one maximal strong clan $b$ in its coarsest quotient, that sees all the remaining maximal strong clans in the same way than the vertex to be added, the case $2.(a)$, we can see it in the center of Figure~\ref{fg:ExampDecomp_3_2a}. Thus, the new vertex is added to it, the resulting clan decomposition tree is shown in the right of Figure~\ref{fg:ExampDecomp_3_2a}.

\begin{figure}[ht]
\centering
\includegraphics[width=.9in]{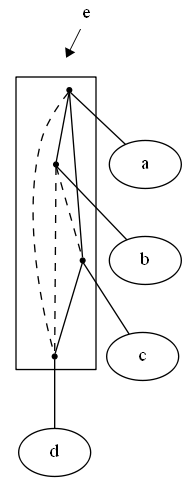}
\centering
\includegraphics[width=1.2in]{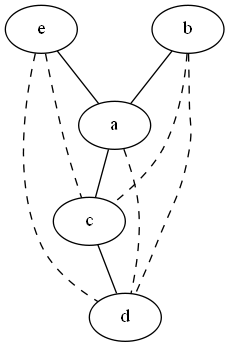}
\centering
\includegraphics[width=1.8in]{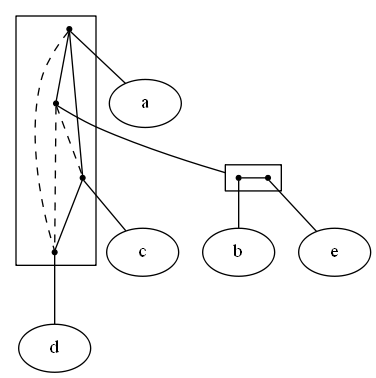}
\caption{Applying 2.(a) from Algorithm~\ref{alg:accordtype}}
\label{fg:ExampDecomp_3_2a}
\end{figure}

When we add $f$ to the primitive clan on the root of the current decomposition, left of Figure~\ref{fg:ExampDecomp_4_2c}, we find $f$ cannot see one of the maximal strong clans, the clan conform by $b$ and $e$, as we have a primitive clan we are in case $2.(c)$, and the elements of the clan are split, right of Figure~\ref{fg:ExampDecomp_4_2c}.

\begin{figure}[ht]
\centering
\includegraphics[width=1.65in]{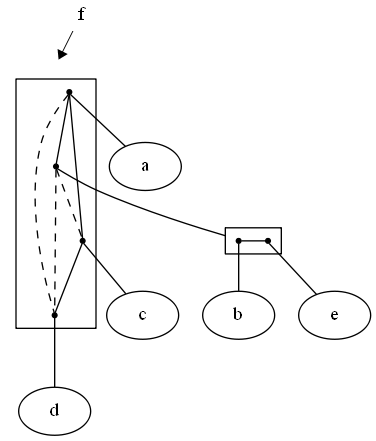}
\centering
\includegraphics[width=1.2in]{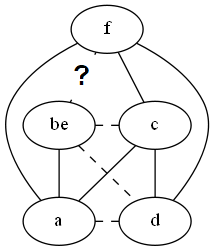}
\centering
\includegraphics[width=1in]{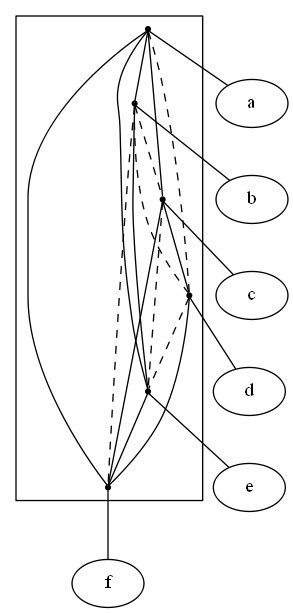}
\caption{Applying 2.(c) from Algorithm~\ref{alg:accordtype}}
\label{fg:ExampDecomp_4_2c}
\end{figure}

In the next step, we add $g$ to the primitive root of this decomposition tree, and we find that $g$ sees all of the maximal strong clans in the coarsest quotient of the root clan in the same way, case $2.(b)$, center of Figure~\ref{fg:ExampDecomp_5_2b}. Thus, this root clan and the new vertex will be in the coarsest quotient of a new complete clan, right of Figure~\ref{fg:ExampDecomp_5_2b}.        

\begin{figure}[h]
\centering
\includegraphics[width=1in]{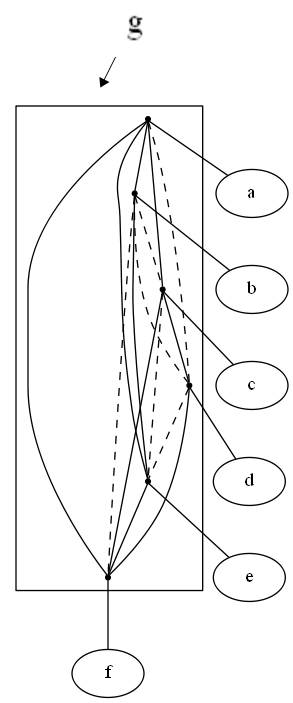}
\centering
\includegraphics[width=1.65in]{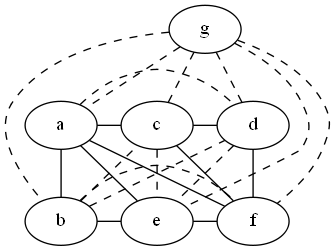}
\centering
\includegraphics[width=1.2in]{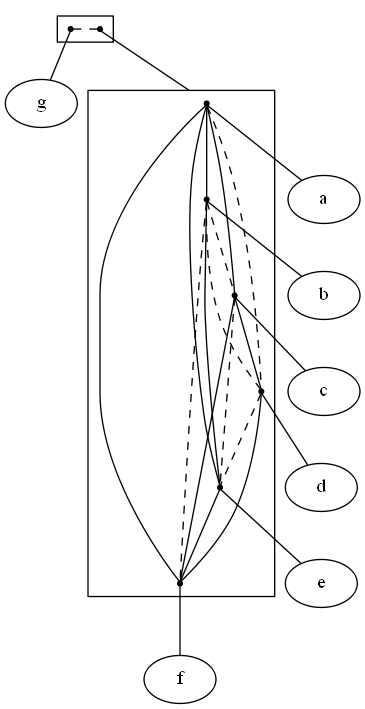}
\caption{Applying 2.(b) from Algorithm~\ref{alg:accordtype}}
\label{fg:ExampDecomp_5_2b}
\end{figure}

When we add $h$ to the complete clan in the root of the current decomposition tree, left of Figure~\ref{fg:ExampDecomp_6_1a}, we find that $h$ sees in the same way than the color of the clan to all the maximal strong clans in the coarsest quotient of the root clan, the case $1.(a)$, center of Figure~\ref{fg:ExampDecomp_6_1a}. Thus, $h$ is added as another maximal strong clan to the coarsest quotient of the complete clan root, right of Figure~\ref{fg:ExampDecomp_6_1a}. 

\begin{figure}[h]
\centering
\includegraphics[width=1.5in]{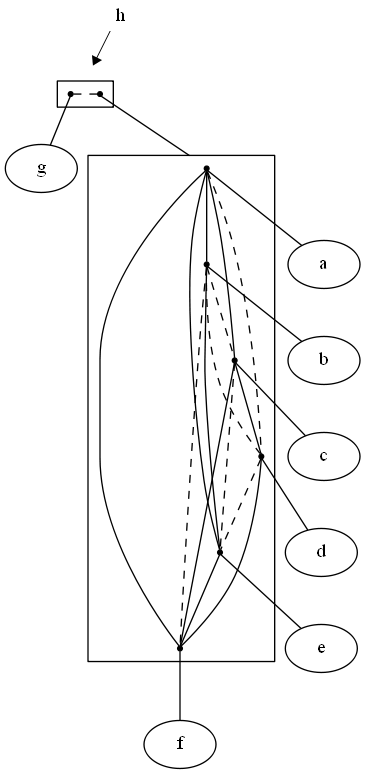}
\centering
\includegraphics[width=1.3in]{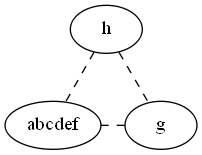}
\centering
\includegraphics[width=1.2in]{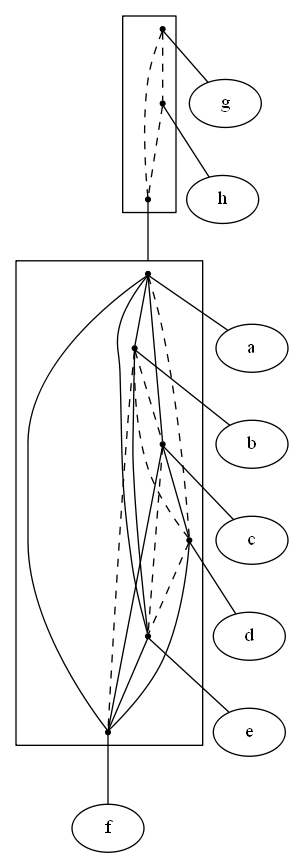}
\caption{Applying 1.(a) from Algorithm~\ref{alg:accordtype}}
\label{fg:ExampDecomp_6_1a}
\end{figure}

Finally, to add $i$ produces a superior clan, since $i$ sees in the same way all the maximal strong clans of the current root complete clan but different than the color of the clan, the case $1.(c)$, center of Figure~\ref{fg:ExampDecomp_7_1c}. Thus, this root clan and the new vertex will be in the coarsest quotient of a new complete clan, the resulting clan decomposition tree is shown in the right of Figure~\ref{fg:ExampDecomp_7_1c}. 

\begin{figure}[ht]
\centering
\includegraphics[width=1.2in]{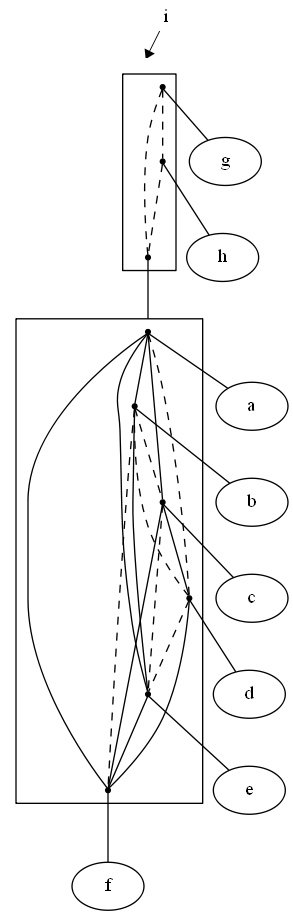}
\centering
\includegraphics[width=1.5in]{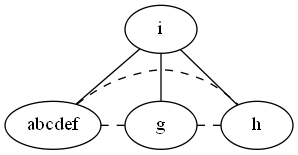}
\centering
\includegraphics[width=1.6in]{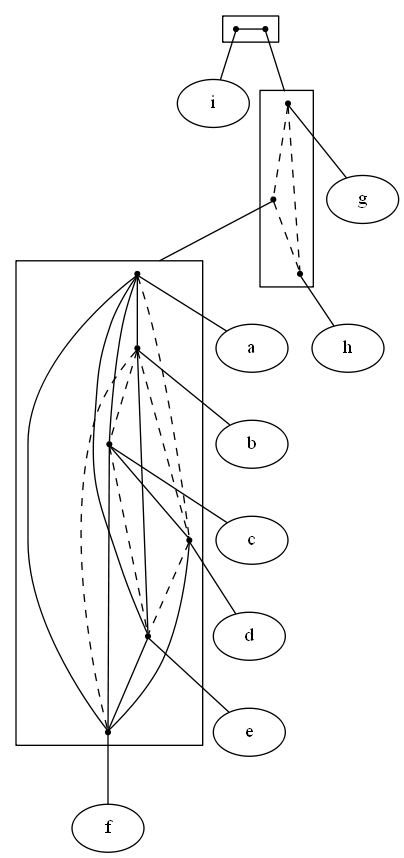}
\caption{Applying 1.(c) from Algorithm~\ref{alg:accordtype}}
\label{fg:ExampDecomp_7_1c}
\end{figure}
\end{example}

\begin{figure}[h]
\centering
\includegraphics[height=0.6\textheight]{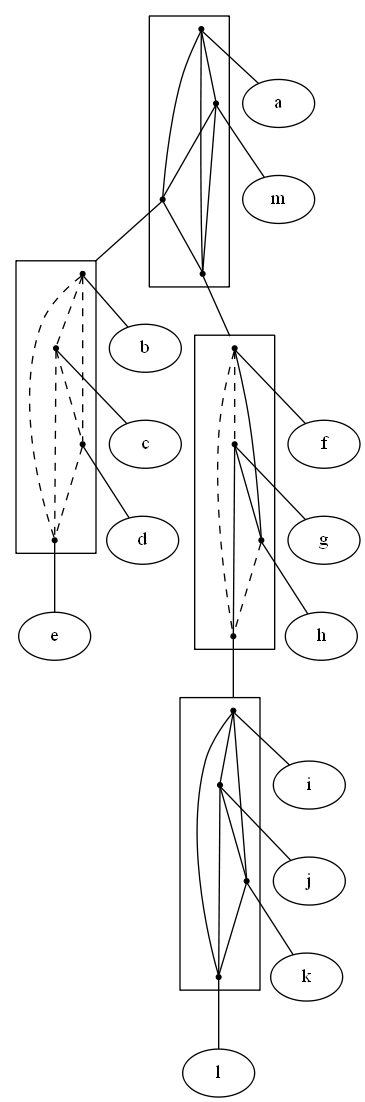}
\centering
\includegraphics[height=0.45\textheight]{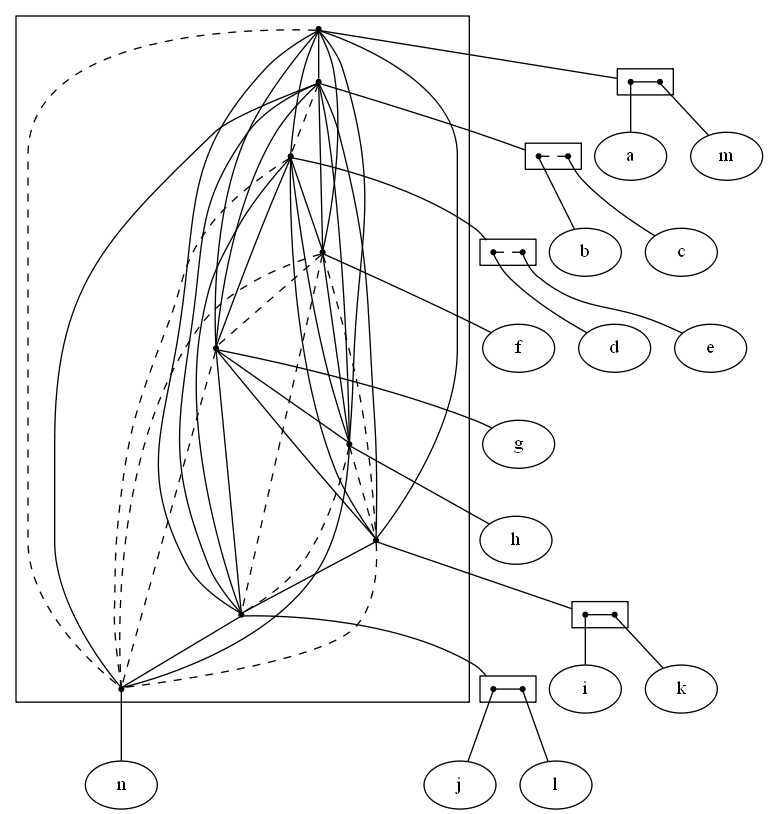}
\caption{Example~\ref{ex:split} to illustrate the use of Theorem~\ref{th:split}}
\label{fg:split}
\end{figure}

\begin{example}
\label{ex:split}
As an example to show the use of split, assume we add $n$ to the decomposition shown in the left of Figure~\ref{fg:split}, such that $n$ 
is connected by solid lines just with $b$, $c$, $h$, $j$ and $l$.

Following the algorithm we will give the maximal strong clans in the coarsest quotient of the current vertices plus the new vertex $n$. 
 
Since the coarsest quotient of the root clan is complete those nodes seen in the same way by $n$ will conform a maximal strong clan, in this case we get the clan $\{a,m\}$.
And those nodes that are not visible from $n$ are split, thus we split the clan $\{b,c,d,e\}$ and the clan $\{f,g,h,\{i,j,k,l\}\}$.
When we split the clan $\{b,c,d,e\}$ we have again it is a complete clan, thus following the previous constrains we get $\{b,c\}$ and $\{d,e\}$ as maximal strong clans.  

As the clan $\{f,g,h,\{i,j,k,l\}\}$ is a primitive clan, the clans into its coarsest quotient that are seen from $n$ continue being maximal strong clans in the new decomposition, while those clans that are not seen by $n$ are split. In this case we continue having $f$, $g$ and $h$ as maximal strong clans, and we split the clan $\{i,j,k,l\}$ since it is not possible to determine an edge to it from $n$. 
Splitting the clan $\{i,j,k,l\}$, as it is a complete clan, we get $\{j,l\}$ and $\{i,k\}$ as maximal strong clans.

The resulting decomposition is shown in the right of Figure~\ref{fg:split}.
 \end{example}

\subsection{Comparative with Other Algorithms}
In this section, we compare our algorithm with a couple of existing algorithms,~\cite{DBLP:journals/dm/McConnellS99} and~\cite{DBLP:journals/algorithmica/McConnell95}:
those that we find conceptually closest.
We refrain from developing a larger perspective as it would be
redundant with an existing, very good survey~\cite{survey}.
We stick to the same terminology of the previous sections
(e.g.,~primitive and complete modules) but warn the reader 
that the sources may employ different nomenclature. 

The algorithm in~\cite{DBLP:journals/dm/McConnellS99} is applied on undirected graphs and it is not incremental. The algorithm starts by constructing a tree that is modified until reaching the modular decomposition tree, going through three kinds of trees.

In the first kind of tree the internal nodes are labeled primitive or complete
by recognizing the $P_4$ induced. If the original graph does not induce any $P_4$,
the initial graph will have just complete internal nodes.

To get the second kind of tree, only the complete nodes of the previous kind of tree will be analyzed in order to find modules within the complete nodes. To reach the third kind of tree, modules that are not strong are removed.

\begin{example}
\label{ex:comp1}

From the left Figure~\ref{fg:diagram0} we get $\{f,h,g,abcde\}$ as induced $P_4$, while the graph conformed by $\{a, b, c, d, e\}$ does not induce $P_4$. In this way, the first kind of tree is a tree with two internal nodes, one labeled as primitive and the other one labeled as complete, right of Figure~\ref{fg:diagram0}. 

The next step is to analyze just the complete modules looking for more modules (strong or not), those only could be complete since all primitive ones were obtained in the previous step. We get, among other modules, $\{abc\}$ as complete strong module and in turn $\{a,b,c\}$ are also modules of it, the process to get the second kind of tree is shown in Figure~\ref{fg:diagramM1M2}.

The last step is to prune the previous tree, removing those modules that are not strong modules. The final decomposition is shown in Figure~\ref{fg:mdspinrad}.
Thus we illustrate the different modus operandi with respect to our algorithm.



\begin{figure}
\centering
\includegraphics[height=0.2\textheight]{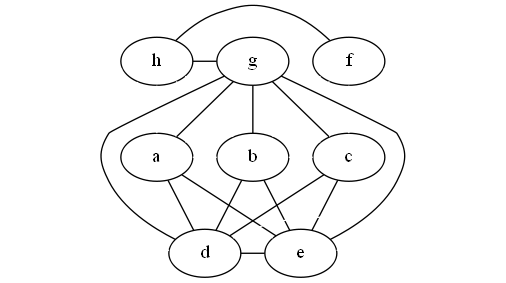}
\centering
\includegraphics[height=0.16\textheight]{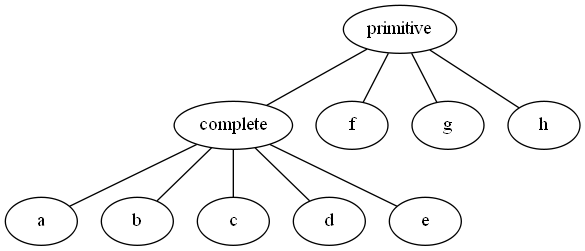}
\caption{Example~\ref{ex:comp1}, initial graph and the first kind of tree}
\label{fg:diagram0}
\end{figure}	

\begin{figure}
\centering
\includegraphics[height=0.16\textheight]{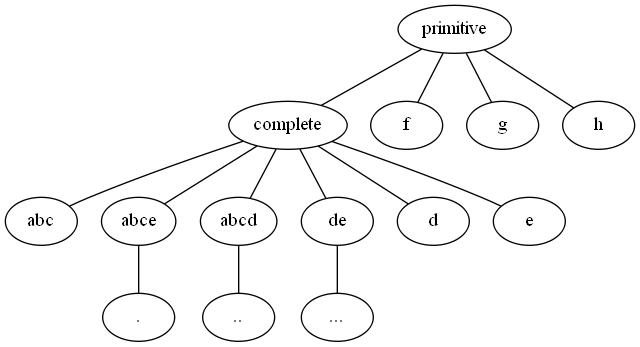}
\centering
\includegraphics[height=0.16\textheight]{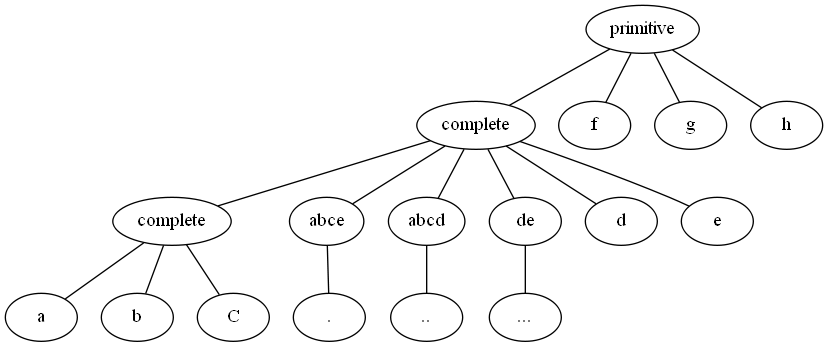}
\caption{Example~\ref{ex:comp1}, process to get the second kind of tree.}
\label{fg:diagramM1M2}
\end{figure}

\begin{figure}
\centering
\includegraphics[height=0.25\textheight]{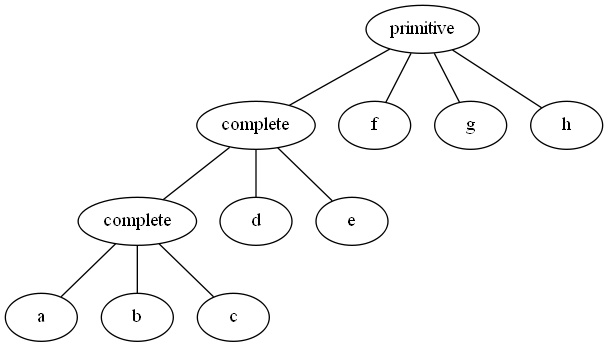}
\caption{Graph of Example~\ref{ex:comp1}, modular decomposition tree.}
\label{fg:mdspinrad}
\end{figure}		

\end{example}

The algorithm in~\cite{DBLP:journals/algorithmica/McConnell95} is applied on all 2-structures, not only symmetric; to simplify the comparison,
since we work only on symmetric 2-structures, we will avoid those parts referring to linear clans. 
It is an incremental algorithm that generalizes the elements of the Muller and Spinrad algorithm for decomposition of undirected graphs~\cite{DBLP:journals/jacm/MullerS89}. Again, despite the fact that this algorithm uses different nomenclature, 
we will try to explain it using our terminology.

The algorithm works with two trees. One of them keeps the clan decomposition of the graph, each internal node of the tree represents a clan but its internal 2-structure is not shown.
The other tree stores node information and information between the nodes by split labels, we will talk a little more about them later.

\begin{example}
\label{ex:compMS}
Assume we add the vertex $e$ to the coarsest quotient graph shown in Figure~\ref{fg:diagram}. 
This new vertex does not see some of the clans in it, $F$ and $G$, and the rest of the elements are seen different than the color of the edges of the coarsest quotient graph, say red (dotted) to $a, b$ and blue (dashed) to $c, d$, right of Figure~\ref{fg:diagram}. Assume $G$ is a primitive clan conform by $g_1, g_2, g_3$ and $F$ is a complete clan conform by $f_1, f_2$, whose coarsest quotients are shown in Figure~\ref{fg:internal}; 
%
besides,                                          
$e$ is connected by different kind of edges with their elements, 
as shown in Figure~\ref{fg:seen}.
 
 \begin{figure}
\centering
\includegraphics[height=0.25\textheight]{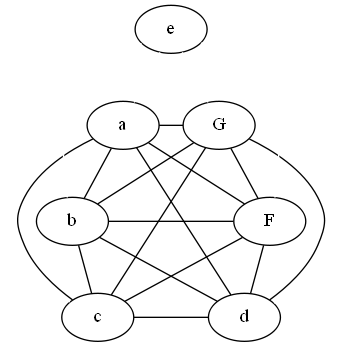}
\includegraphics[height=0.25\textheight]{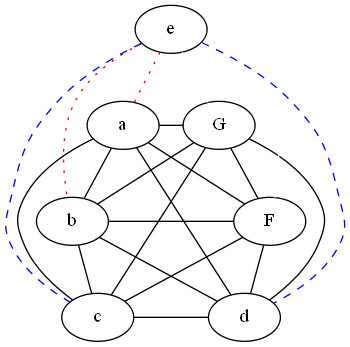}
\caption{Example~\ref{ex:compMS}: 2-structure to add node $e$.}
\label{fg:diagram}
\end{figure}	

\begin{figure}
\centering
\includegraphics[height=0.04\textheight]{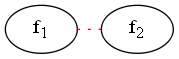}
\includegraphics[height=.1\textheight]{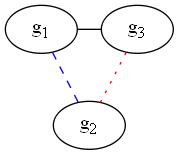}
\caption{Example~\ref{ex:compMS}: Coarsest quotient graphs of $F$ and $G$}
\label{fg:internal}
\end{figure}	

\begin{figure}
\centering
\includegraphics[height=0.1\textheight]{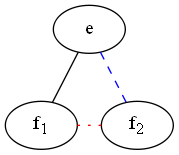}
\includegraphics[height=0.14\textheight]{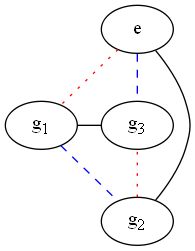}
\caption{Example~\ref{ex:compMS}: How $e$ sees the clans into $F$ and $G$ }
\label{fg:seen}
\end{figure}	

The modular decomposition before to add $e$ is shown 
in Figure~\ref{fg:MD1}.
The nodes are filled by white if there is an unambiguous color to connect the node with $e$, and gray otherwise (``not seen'').

\begin{figure}
\centering
\includegraphics[height=0.14\textheight]{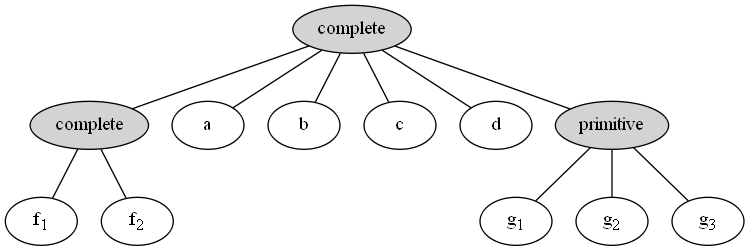}
\caption{Initial tree decomposition and tree decomposition according $e$ sees its nodes}
\label{fg:MD1}
\end{figure}	

\begin{figure}
\centering
\includegraphics[height=0.09\textheight]{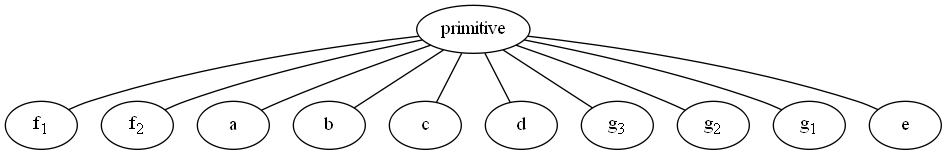}
\caption{Final tree decomposition}
\label{fg:final}
\end{figure}

The algorithm ask for the type of the node that is not seen by $e$.
If the node is primitive its children are the same plus the new vertex.
Else (the node is complete), its children will be those nodes that are not connected with $e$ by the same kind of edges than the edges of the node.

The other tree stores each visible clan of the previously described tree.
If the node is complete its children are grouped according to how they are seen from the new vertex ($e$, in this example). 
The edges are labeled by a split label: $w$ is a split label for the edges $X$,$Y$ if $(X,w) \neq (Y,w)$. 
In this way, this tree works complementing the information about the children of the nodes in the decomposition tree.
The final tree decomposition is shown in Figure~\ref{fg:final}.

%
\end{example}

The main difference with Algorithm~\ref{alg:accordtype} is that, instead of building this tree and doing the labeling process, our algorithm uses the Union-Find structure: each time a new clan is generated, its edges to the other clans (sigleton or not) are added according to their equivalence classes if they exist.

\section{Discussion and perspectives}

Along this paper, we have shown, first, that the known notion
of modular decomposition of a graph can be understood, in a quite natural
way from a perspective of data analysis, as a variant of 
closure space visualization; then, that this process can be applied to
transactional datasets via a known logical construction, namely the
Gaifman graph; and, also, that both the theoretical connection and
the practical applicability of the decomposition process
can be generalized to quantitatively
enabled variants of Gaifman graphs through a known generalization
of modular decompositions, namely the clan decomposition of 2-structures.
We have included a number of developments that relate the taxonomy
of modules to the corresponding local structure of the closure
space, with algorithms to construct the adequate 2-structure to
understand which equivalence classes are relevant for the 
decompositions and which ones are not; and we have described
an incremental algorithm, which extends existing ones, that we
are currently using in our open-source software 
tools,~
\url{https://github.com/MelyPic/Gaifman-graphs\_Mod},
that we use in order to compute clan decompositions. 
There is ample room to study improvements to this algorithm 
and, as mentioned above, the possibilities of using our 
theoretical results to obtain additional, alternative, 
hopefully better algorithms.

The avenues for further research are many and wide. First and
foremost, additional examples of the practical relevance of
this approach are convenient; so far, we can offer some 
interesting applications
in \cite{DBLP:conf/ida/BalcazarPR18} and \cite{CBMS}, but
we hope to add further, equally convincing case studies down the road.

Along this practical line, we have found many cases where the
obtained graphs and decompositions are somewhat too large or
complex to provide intuition through diagrams.
We have
proposed a few simple strategies to encompass complex
substructures upon visualization in 
\cite{DBLP:conf/ida/BalcazarPR18}
(our ``Others'' nodes in the examples, for instance) 
but a more systematic study of the ways in which
visual diagrams
can become helpful is necessary; 
we believe that the answers will come from some 
notion of interactive data analysis process: 
hence our insistence that the major decomposition
algorithms to be used should be incremental.

Admittedly, our proposals for enhancing Gaifman graphs with
quantitative information can be considered simple or even
na\"\i{}ve, and must be subject to further testing on
practical cases and to comparison with additional alternatives
that could be designed in the future. For instance, one
may focus on the set of integers arising as multiplicities of
each of all the pairs of vertices in our quantitative 
Gaifman graph, and consider this set of integers as a 
single-dimensional dataset itself; then, on it, 
bring to bear existing unsupervised discretization 
methods in order to split the edges into their
corresponding equivalence classes: the options explored here
amount to sticking to the ``equal length bins'' discretization
(either on the original data or after a logarithmic scaling),
and quite a few additional options exist.

Also, many other tunings can be applied to the Gaifman graph before
applying the decomposition procedure. For instance, we have
started in \cite{WomEncourage} the exploration of the case
where the equivalence class of the 2-structure edge $(x,y)$
depends on the distance between $x$ and $y$ along a shortest
path of the original Gaifman graph. 
Other similar parameters such as connectivity (number
of disjoint paths, that one can relate to Menger's theorem)
could be applied as well.

The multirelational potential is also to be developed. Of course,
if we are given a dataset consisting of several tables, it is
a simple matter to apply directly our approach as, indeed,
Gaifman graphs were defined from the start so as to apply
to any relational structure with possibly several tables:
a Gaifman edge would join $x$ and $y$ whenever they appear
together in some tuple of some table.
However, initial explorations in \cite{DBLP:conf/ida/BalcazarPR18}
suggest that this na\"\i{}ve application will fall short of
providing good results, because a crucial notion 
in multirelational data, namely foreign keys, is being
ignored. How to take foreign keys into account at the time of
constructing the Gaifman graph, in a way that provides
sensible results in practice, is a question that needs
careful exploration, both conceptually and in terms of
efficiency (e.g.~we could denormalize into a single large
table the data through the foreign keys, but this looks
like an inefficient implementation even if the process
turns out to be practically applicable at all). 
We believe that a generalization of the ``shortest path'' 
variant alluded to above could lead to a working approach.

Further theoretical issues remain open. To mention just
two of them: in our scheme according to Figure~\ref{fg:scheme},
would there be a way of telling whether a given set
of implications will be amenable to the graph reconstruction
procedure?, and, regarding the clan taxonomy given the
closure space, how to get further information about the
internals of each clan in a way similar to the facts we
have been able to establish about modules?

Finally, it is well-known that, in data analysis tasks,
categorical concepts (such as implications) benefit from
a relaxation (such as partial implications or association
rules) allowing for exceptions, whether they come from
varied inputs or even from material errors in coding or 
transmission. Likewise, we could relax through allowing
exceptions the notions of module and clan. The concept we would 
end up with seems to us very close to (a recursive form
of) the notion of ``blockmodeling'' employed in social
network analysis \cite{BlockModelingBook}; this is
again a large area where, to start with, a clarification 
endeavor will be effortful but necessary.

\section*{Acknowledgements}
We are grateful to Laura Rodr\'\i{}guez-Navas for her help in
many ways, to the Hospital de la Santa Creu i Sant Pau for allowing 
our research group to employ their diagnostic data in our research, 
and to Ricard Gavald\`a who provided the data in such a clean form.

This work was partially supported by 
{European Research Council} 
 under the European Union's Horizon 2020 
   research and innovation programme,
 grant agreement
ERC-2014-CoG 648276 (AUTAR)
, by~
{Ministerio de Economia, Industria y Competitividad}
, project
{TIN2017-89244-R}%
, and by
{Conacyt (M\'exico).} 
We acknowledge unfunded
recognition by
{AGAUR (Generalitat de Catalunya)} 
as research group 
{2017SGR-856 (MACDA)}.

\bibliographystyle{splncs03}

\end{document}